\documentclass[11pt,catalan]{article}
\usepackage{graphicx}
\usepackage{amsmath}
\pdfoutput=1
\usepackage{amsfonts}
\usepackage{amssymb}
\usepackage{bm}

\usepackage[english]{babel}
\usepackage[latin1]{inputenc}
\usepackage{times}
\usepackage[T1]{fontenc}

\usepackage{cite}

\newcommand{\RR}{\ensuremath{ \mathbb{R}}}
\newcommand{\ZZ}{\ensuremath{ \mathbb{Z}}}

\newcommand{\ri}{{\rm i} }

\newcommand{\sign}{{\rm sign}}

\newcommand{\D}{{\rm d}}

\DeclareMathOperator{\G}{\mathcal{G}}
\DeclareMathOperator{\E}{\mathcal{E}}
\DeclareMathOperator{\T}{\mathcal{T}}
\DeclareMathOperator{\intR}{\int_{\mathbb{R}}}

\DeclareMathOperator{\intRRR}{\int_{\mathbb{R}^3}}

\newtheorem{proposition}{Proposition}

\newcounter{llista}

\begin{document}
\title{Non-local Lagrangian fields: Noether's theorem and Hamiltonian formalism}
\author{Carlos Heredia\thanks{e-mail address: carlosherediapimienta@gmail.com}\,\,\, and Josep Llosa\thanks{e-mail address: pitu.llosa@ub.edu}\\
Facultat de F\'{\i}sica (FQA and ICC) \\ Universitat de Barcelona, Diagonal 645, 08028 Barcelona, Catalonia, Spain }
\maketitle

\begin{abstract}
This article aims to study non-local Lagrangians with an infinite number of degrees of freedom.  We obtain an extension of Noether's theorem and Noether's identities for such Lagrangians. We then set up a Hamiltonian formalism for them. In addition, we show that $n$-order local Lagrangians can be treated as a particular case and the standard results can be recovered. Finally, this formalism is applied to the case of $p$-adic open string field. 
\end{abstract}

\section{Introduction \label{S1}}

One of the most frequently emerging features in quantum gravity models is non-locality. In string theory, for instance, non-locality is displayed in its interactions, characterized by its infinite derivative structure \cite{WITTEN1986,Seiberg1999}. In a more visual way, the interactions are not point-wise but are given in a specific finite region. A similar idea occurs in the case of loop quantum gravity \cite{Ashtekar2013} or effective models of string theory such as $p$-adic strings \cite{Volovich1987,BREKKE1988}.

At the classical level, non-local gravity models -- inspired by the ultraviolet (UV) finiteness of string theory, see for instance \cite{Biswas2006} --- are being proposed to solve both cosmological and black hole singularities. There is an essential improvement in the UV regime by adding infinite derivatives to the Lagrangian without introducing new degrees of freedom \cite{Biswas2012}. These non-local models of gravity are called Infinite Derivative Theories of Gravity (IDGs), and their results are quite promising; for instance, they can show the regularisation of the gravitational potential $1/r$ of pointlike sources at the linearised level \cite{Edholm2016}, as well as other sorts of sources \cite{Boos2021,Kol2020,Kol2021,Buoninfante2018_2,FrolovValeri2015,Dengiz2020,Kumar2021}. Likewise, other non-local gravity models are also being used to explain the cosmic expansion of the Universe \cite{Capozziello2021}. It was shown that the $1/\Box$ operator applied on the R-curvature scalar results in an accelerated expansion of the Universe without relying on a contribution from dark energy \cite{Deser2007}.

All these models mentioned contain non-locality both in space and time. Spatial non-locality might be considered a mere curiosity presented by the theory; however, temporal non-locality is a very problematic feature in the sense of the initial value problem and the preservation of causality. However, recent studies \cite{Barnaby2008,CALCAGNI2008,Gorka2012,Erbin2022} show that the initial value problem might be well-posed even though infinite derivatives or integrodifferential equations are involved.  Likewise, it is shown in \cite{Gorka2011,Gorka2012_2,Gorka2013,Bravo2019} that the existence of solutions for elliptic partial differential equations containing infinitely many derivatives might be slightly more manageable to provide. 

In the 1990s and 2000s, a Hamiltonian formalism for non-local Lagrangians was developed \cite{Jaen1987,Llosa1994} and was known as  (1+1)-dimensional Hamiltonian formalism \cite{Gomis2001}. The main idea of this formalism was to rewrite the non-local Lagrangian into a local-in-time one by using an extra dimension and thus be able to formulate the Hamiltonian formalism in this equivalent theory. Later, this formalism was applied in \cite{Gomis2001_2,SUYKENS2009,Kolar2020}, among other cases. Unfortunately, as Ferialdi et al. \cite{Ferialdi2012} correctly pointed out, this approach is lacking in considering non-local Lagrangians that explicitly depend on time. However, in a recent paper \cite{Heredia2},  this formalism was significantly improved by extending Noether's theorem for non-local Lagrangians mechanics,  i. e. a finite number of degrees of freedom. This extension allowed the use of a conserved quantity  to infer a suitable definition for the Legendre transform and thus avoid the extra non-physical dimension. Furthermore, the conserved quantity was presented in a closed form --- i.e. with the infinite series that appears when dealing with infinite-order Lagrangians summed---, and  the deficiency of considering explicitly time-dependent non-local Lagrangians was addressed.

The present work aims to adapt the latter results \cite{Heredia2} to non-local Lagrangian fields, considering all the peculiarities of field theories with respect to mechanics. In Section 2, we present the functional form of non-local field Lagrangians, which may explicitly depend on the spacetime point. We then raise the non-local variational problem and derive the Lagrange field equations. It happens that any Lagrangian is the total 4-divergence of a non-local current but this does not imply that the Lagrange equations vanish identically. We find what extra conditions the non-local current must satisfy for the Lagrange equations to vanish identically.  

In Section 3, we study Noether's symmetries, including the case that the Lagrangian explicitly depends on the spacetime point, and we find the conserved currents associated to symmetry finite Lie groups ---first Noether's theorem---  . We then concretize to Poincar\'e invariant field theories and derive both the energy-momentum and angular momentum tensors. Then, by inspecting the expression of the energy density, we can guess the form of the Legendre transformation that, in Section 4, allows us to set up a Hamiltonian formalism for the non-local Lagrangian field theory and derive a precise expression for the Hamiltonian and the symplectic form. Finally, in Section 5, we apply all these tools to the $p$-adic open string. By using a perturbative solution, we obtain the Hamiltonian, a set of canonical coordinates and, by canonical quantization, we set up a quantum theory. In addition, we calculate each of the components of the Belinfante-Rosenfeld tensor in closed form and we obtain the total linear momentum and the pressure exerted on a spherical surface. 

\section{Non-local Lagrangian theories \label{S2}}
Consider the action integral
\begin{equation} \label{A0}
  S = \int_{\RR^4} \D x\,\mathcal{L}([\phi^A],x)  \,, 
\end{equation}
where the Lagrangian density $\mathcal{L}$ depends on all the values $\phi^A(z)$, $\,A=1 \ldots m\,$, of the field variables at points $z$ other than $x$. This fact is why we refer to it as non-local. Likewise, we take  $x\in \RR^4$  for concreteness; however, the following also holds for any number of dimensions.

The class of all possible fields, whether or not they meet the field equations ---{\em off shell}---, makes up the {\em kinematic space} $\mathcal{K}$. This space is the subspace of all smooth functions $\mathcal{C}^\infty(\RR^4;\RR^m) $ such that $\mathcal{L}([\phi^A],x)\,$ is locally summable. For Lagrangians depending explicitly on the point\footnote{As in the case of $p$-adic string \cite{Barnaby2011}.} $x^b$, we have to resort to the {\em extended kinematic space}, $\mathcal{K}^\prime= \mathcal{K} \times\RR^4$. 

\vspace{0.1cm}
The non-local Lagrangian density is a real-valued functional  
$$ (\phi^A,x^b) \in \mathcal{K}^\prime \longrightarrow \mathcal{L}(\phi^A,x^b) \in \RR \,,$$
and it may depend on all the values $\phi^A(z)\,, \;\; z\in\RR^4\,$. To make the notation lighter, we will write $\mathcal{L}(\phi,x)$, where the functional dependence is understood although the square bracket does not emphasize it, as is usually done in most textbooks. Moreover, we also omit the superindices both in the field variables and the point coordinates, unless the context makes it necessary, e. g. in Section \ref{Noe}.

\vspace{0.1cm}
The function $\phi(z)$ contains all information about the evolution in $\mathcal{K}^\prime$. Given $ y\in \RR^4$, we define the spacetime translation
\begin{equation}  \label{L1} 
  (\phi,x) \stackrel{T_y}{\longrightarrow} (T_y \phi, x+ y) \,,\qquad {\rm where} \qquad T_y \phi(z) = \phi( y+z)\;,
\end{equation}
which has the obvious additive property $\; T_{ y_1}\circ T_{ y_2} = T_{ y_1+ y_2}\,$.
We will refer to the subset \\
$\;\{(T_y \phi, x+ y),\; \, y\in \RR \}\subset \mathcal{K}^\prime\;$ as the {\em field trajectory} starting at $(\phi,x)\,$.

\vspace{0.1cm}
The action integral (\ref{A0}) is currently understood as the functional on $\mathcal{K}$ 
\begin{equation}   \label{A1o}
 S(\phi) := \int_{\RR^4} \D  y\, \mathcal{L}\left(T_y\phi,  y \right) \,. 
\end{equation}
It may be divergent because we need an unbounded integration domain to abide by  the fact that the Lagrangian density $\mathcal{L}$ depends on all the values $\phi(z)\,$. An alternative and more consistent formulation is introducing the 1-parameter family of finite action integrals 
\begin{equation} \label{A1}
  S(\phi,R) = \int_{|y|<R} \D y\,\mathcal{L}(T_y\phi,y)  \,, \qquad R\in \RR^+ \,,
\end{equation}
where $\,|y|= \sqrt{\sum_{j=1}^4 (y^j)^2 }\;$ is the Euclidean length. 
Then the variational principle reads
\begin{equation} \label{A2}
\lim_{R\rightarrow\infty} \delta S(\phi,R) \equiv \lim_{R\rightarrow\infty}  \int_{|y|<R} \D y\,\int_{\RR^4} \D z\,\frac{\delta \mathcal{L}\left(T_y \phi,  y \right)}{\delta \phi(z)} \,{\delta \phi(z)} = 0\,,  
\end{equation} 
for all variations $\,\delta \phi(z) \,$ with compact support, and the Lagrange equations are
\begin{equation}  \label{L2o} 
\psi(\phi,z) = 0 \,, \quad {\rm with} \quad \psi(\phi,z) := \int_{\RR^4} \D  y\,\lambda(\phi, y,z) \quad {\rm and} \quad
\lambda(\phi, y,z) := \frac{\delta \mathcal{L}\left(T_y\phi,  y\right)}{\delta \phi(z)} \:.
\end{equation}
The {\em dynamic fields} ---on shell--- are those $\phi$ fulfilling this equation.

\vspace{0.1cm}
So far, the variational principle formulation has been limited to trajectories $\,(T_y\phi,  y)\,$  initiating at $(\phi,0)$. As we are interested in Lagrangians that may explicitly depend on the point, we need to extend this formulation to abide trajectories starting at any $\,(\phi,x) \in\mathcal{K}^\prime\,$. Such a trajectory is nothing but the one starting at $(T_{-x}\phi,0)$ but advanced an amount $x$, that is
\begin{equation} \label{L1a}
 (T_y\phi, x+ y) =  (T_{ y^\prime} \tilde\phi,  y^\prime)\,, \qquad {\rm with}
\qquad  y^\prime = x+ y \quad {\rm and} \quad  \tilde\phi= T_{-x} \phi \:.
\end{equation}
Hence, the Lagrange equation for the dynamic trajectory initiating at $(\phi,x)$ is
\begin{equation}  \label{L2} 
\Psi(\phi,x,z) = 0 \,, \quad {\rm where} \quad \Psi(\phi,x,z) := \psi(T_{-x}\phi,z+x) 
\end{equation}
or $\quad\Psi(\phi,x,z) := \int_{\RR^4} \D  y\,\Lambda(\phi,x, y,z)\quad$ with 
\begin{equation}  \label{L2z} 
\Lambda(\phi,x, y,z) := \lambda(T_{-x}\phi, y+x,z+x) = \frac{\delta \mathcal{L}(T_y\phi,x+ y)}{\delta\phi(z)}\;.
\end{equation}
The following property easily follows from the definition:
\begin{equation}  \label{L2y} 
\Lambda(T_u\phi,x+ u,y,z) = \lambda(T_{-x}\phi,y+x+ u,z+x+ u) = \Lambda(\phi,x,y+ u,z+ u)\;,
\end{equation}
which will be useful later.

\subsection{Local theories as a particular case \label{S1.1}}
Let us see how a standard Lagrangian $L\left( \phi, \ldots \phi_{|{b_1 \ldots b_k}},x\right)$, which depends on the field derivatives up to the order $k$, fits in the formalism developed so far --- the ``stroke'' means ``partial derivative''---. The standard action integral $\;\;\int \D x\,L(\phi(x),\ldots \phi_{|b_1\ldots b_k}(x),x)\;$ has the form (\ref{A1}) provided that we take
\begin{equation}  \label{A2a}
\mathcal{L}(T_y\phi, y) := L(\phi( y),\ldots \phi_{|b_1\ldots b_k}( y), y)\:.
\end{equation}
Whence it follows from (\ref{L2o}) that
\begin{equation}  \label{L2a} 
\lambda(\phi, y, z) = \frac{\delta \mathcal{L}\left(T_y\phi,  y\right)}{\delta \phi( z)} = \sum_{j=0}^k \left(\frac{\partial L}{\partial \phi_{|c_1\ldots c_j}}\right)_{(\phi( y), \ldots, \phi_{b_1\ldots b_k}( y), y)}\,(-1)^j\delta_{|c_1\ldots c_j}( z- y)  \,,
\end{equation}
where we have included that 
$$\phi_{c_1\ldots c_j}( y) = (-1)^j\,\int_{\RR^4} \D z\,\phi( z)\,\delta_{|c_1\ldots c_j}( z- y)\,.$$
Substituting (\ref{L2a}) in (\ref{L2z}), we finally arrive at
\begin{equation}  \label{L2b} 
 \Psi(\phi,x, z) \equiv \sum_{j=0}^k (-1)^j\,\frac{\partial^j}{\partial  z^{c_1} \ldots\partial z^{c_j}} \left(\frac{\partial L}{\partial \phi_{|c_1\ldots c_j}}\right)_{(\phi( z), \ldots ,\phi_{b_1\ldots b_k}( z),x+ z)} = 0\;,
\end{equation}
which is the  {\em Euler-Ostrogradski equations} \cite{Ostrogradski}.

\subsection{The Lagrange equations for a total divergence \label{S2.1a}}
A well-known feature of local theories is that, when the Lagrangian density is a total divergence,
\begin{equation}  \label{TD1}
 \mathcal{L}(x) = \partial_b W^b(x) \,,
\end{equation}
then the Lagrange equations vanish identically. 
The non-local case is more nuanced since equation (\ref{TD1}) has always a solution (in fact, infinitely many). Indeed, the general solution is 
$$ W^b(x) = \delta_4^b\,\int_\RR \D \tau\,\left[\theta(\tau)-\theta(\tau-x^4)\right]\,\mathcal{L}(\mathbf{x},\tau) + \partial_c\Omega^{bc}(x) \,,$$ 
where $\,x=(\mathbf{x},\tau)\,$ and $\Omega^{bc}+\Omega^{cb} = 0\,$. 
However, as the solution $\, W^b(\phi,x) \,$ is not necessarily local, it does not imply that the Lagrange equations for any non-local Lagrangian density are identically null.

\vspace{0.1cm}
Let us now search for a sufficient condition on $\, W^b(\phi,x) \,$ for the Lagrangian $\,\partial_b W^b(\phi,x) \,$ to produce null field equations. The family of actions (\ref{A1}) for such a Lagrangian is
$$   S(\phi,R) = \int_{|y|<R} \D y\,\partial_b W^b(T_y\phi,y)  = \int_{|y|=R} \D \Sigma_b(y)\,W^b (T_y \phi,y) $$
where Gauss theorem has been applied and $\,\D \Sigma_b(y)\,$ is the volume element on the hypersphere $\,|y|=R\,$. 

\vspace{0.1cm}
The variational principle (\ref{A2}) yields the field equations
$$ \psi(\phi,z) :=\lim_{R\rightarrow\infty} \frac{\delta S(\phi,R)}{\delta \phi(z)} \equiv 
\lim_{R\rightarrow\infty} \int_{|y|=R} \D \Sigma_b(y)\,\frac{\delta W^b (T_y \phi,y)}{\delta\phi(z)} $$ 
and, as $\, \D \Sigma_b(y)\,$ scales as $|y|^3$, they are identically null provided that 
\begin{equation} \label{TD3}
 \lim_{|y|\rightarrow\infty} \left\{|y|^3\,\frac{\delta W^b (T_y \phi,y)}{\delta \phi(z)} \right\} \equiv 0  
\end{equation}
where the symbol $\equiv$ means that the equalties hold for any $\phi\,$. This condition is obviously met if $ W^b (T_y \phi,y)$ is local, i. e. it depends only on a finite number of derivatives of $\phi$ at $y$.  

\subsection{The Lagrange equations, time evolution and spacetime translations \label{S2.2}}
Obviously, equation (\ref{L2}) is not met by any $(\phi, x) \in\mathcal{K}^\prime\,$. Therefore, the Lagrange equation acts as an implicit equation defining the {\em dynamic space} $\mathcal{D}^\prime$, i.e. the class of all dynamic fields, as a submanifold of $\mathcal{K}^\prime\,$.

\vspace{0.1cm}
In the local regular case, equation (\ref{L2b}) is a partial differential system of order $\,2 k\,$ which usually admits a well-posed Cauchy problem. According to the Cauchy-Kowalewski theorem \cite{john1991}, given a non-characteristic hypersurface $\Sigma$ in $\RR^4$ with normal vector $\,n^b\,$ and $\,2 k\,$ functions, $\; u_j\,,\; j= 0 \ldots 2k-1\,$, on $\Sigma$, there exists a solution $\,\phi(x)\,$ of the PDE (\ref{L2b}) such that
$$ n^{b_1} \ldots n^{b_j} \phi_{|b_1 \ldots b_j}(x) = u_j(x)\,, \qquad j= 0 \ldots 2k-1\,,\qquad \forall x\in\Sigma \,.$$ 

\vspace{0.1cm}
In case that $\Sigma$ is the hyperplane $\,t=x^4 = 0\,$, then $n^b= (0,0,0,1)\,$ and the Cauchy-Kowalevski theorem is the basis for interpreting the Cauchy data $\; u_j(\mathbf{x})\,,\; j= 0 \ldots 2k-1\,$ as ``the state of the field'' at $t=0$, which evolves in time steered by the field equation (\ref{L2b}).

\vspace{0.1cm}
Furthermore, and similarly as the theorems of existence and uniqueness do for systems with a finite number of degrees of freedom, the Cauchy-Kowalevski theorem allows parametrizing each solution $\phi \in \mathcal{D}$ with a well-defined --- although infinite --- set of ``parameters'', namely the Cauchy data.

\vspace{0.1cm}
In contrast, the above interpretation does not hold for a non-local Lagrangian because, as a rule, we do not have an equivalent to the Cauchy-Kowalevski theorem to turn to. For this reason, we  take (\ref{L2}) as an implicit equation or constraint defining $\mathcal{D}^\prime$ as a submanifold of $\mathcal{K}^\prime\,$ that we write as
\begin{equation}  \label{L3} 
 \Psi\left(\phi, x\right) = 0 \qquad \forall \sigma\in\RR^4 \,, \qquad \qquad \mathrm{where}\qquad\qquad \Psi\left(\phi, x\right)_{(\sigma)} := 
\int_{\RR^4} \D  y\,\Lambda(\phi,x, y,\sigma)\,.
\end{equation}
The notation is intended to indicate that $\Psi$ maps $\mathcal{K}^\prime\,$ on the space of smooth functions of $\sigma\in \RR^4$. The dynamic fields are those $(\phi,x)$ that make $\Psi $ null.

\vspace{0.1cm}
The infinitesimal generators of spacetime translations (\ref{L1}) in $\mathcal{K}^\prime$ are the vector fields $\,\mathbf{X}_a\,$, $\;a = 1\ldots 4\,$, that are tangent to the curves $(T_y \phi, x+ y)\,, \;\,  y^b = \varepsilon\,\delta_a^b\,$. Therefore, for a function  $F(\phi,x)$ on $\mathcal{K}^\prime$, we have that
\begin{equation}  \label{L4} 
\mathbf{X}_a F(\phi,x) := \left[\frac{\partial F\left(T_y \phi, x+ y\right)}{\partial \varepsilon} \right]_{\varepsilon=0} \,, \qquad  y^b = \varepsilon\,\delta_a^b\,.
\end{equation}
They are vector fields on $\mathcal{K}^\prime\,$ that, including the chain rule, can be written as
\begin{equation}  \label{L5} 
\mathbf{X}_a := \partial_a  + \int_{\RR^4} \D\sigma\,\phi_{|a}(\sigma)\,\frac{\delta \quad}{\delta \phi(\sigma)} 
\end{equation}
where $\mathbf{X}_4$ is the generator of time evolution and will play a central role in Section \ref{Ham}.

\vspace{0.1cm}
For the particular way in which we have defined equation (\ref{L2}) as an extension of (\ref{L2o}), the constraints (\ref{L3}) are stable under spacetime translations, and therefore the generators $\,\mathbf{X}_a\,$ are tangent to the submanifold $\mathcal{D}^\prime \subset \mathcal{K}^\prime\,$. Indeed, including (\ref{L3}) and (\ref{L2y}), we have that 
$$  \Psi\left(T_y\phi, x+ y \right)_{(\sigma)} = \int_{\RR^4} \D\tau\,\Lambda\left(T_y\phi,x+ y, \tau,\sigma \right) = \int_{\RR^4} \D\tau^\prime\,\Lambda\left(\phi,x, \tau^\prime,\sigma+ y \right) = \Psi\left(\phi,x \right)_{(\sigma+ y)} \,,$$
where (\ref{L2y}) has been included and the replacement $\tau^\prime=\tau+ y$ has been made.
Hence, if $\Psi(\phi,x)=0\,$, then $\Psi\left(T_y\phi, x+ y \right)=0$ as well, and therefore
$$ \mathbf{X}_a \Psi(\phi,x)_{(\sigma)} = \left[\frac{\partial \Psi\left(T_y\phi, x+ y \right)_{(\sigma)}}{\partial \varepsilon} \right]_{\varepsilon=0} = 0 \,, \qquad  y^b = \varepsilon\,\delta_a^b  \,.$$

\section{Noether's theorem  \label{Noe}}
What follows, let us restore the superindex $A$ in the field variable since it will be necessary for non-scalar fields. Consider the infinitesimal transformation
\begin{equation}  \label{A3}
x^{\prime a}(x) = x^a + \delta x^a(x) \,, \qquad \quad \phi^{\prime A}(x) = \phi^A(x) + \delta \phi^A(x) \,.
\end{equation}
The Lagrangian density transforms so that the action integral over any 4-volume is preserved, that is 
$$ S(\mathcal{V})=S^\prime (\mathcal{V}^\prime)\,, \qquad {\rm with} \qquad  S(\mathcal{V}) \equiv \int_{\mathcal{V}} \D x \,\mathcal{L}\left(T_x\phi^A, x\right)  \qquad {\rm and} \qquad  
S^\prime (\mathcal{V}^\prime) \int_{\mathcal{V}^\prime} \D  x^\prime\, \mathcal{L}^\prime\left(T_{ x^\prime}\phi^{\prime A}, x^\prime \right) \,,$$
where ${\mathcal{V}^\prime}$ is the transformed of the spacetime volume $\mathcal{V}\,$.
Therefore
$$ \mathcal{L}^\prime\left(T_{ x^\prime}\phi^{\prime A}, x^\prime \right) = \mathcal{L}\left(T_x\phi^A, x\right) \,\left|\frac{\partial x}{\partial x^\prime}\right| \,.$$

We say that the transformation (\ref{A3}) is a {\em Noether symmetry} if the transformed Lagrangian is the original one plus a total divergence, namely
\begin{equation}  \label{A4a}
  \mathcal{L}^\prime\left(T_{ x^\prime}\phi^{\prime A}, x^\prime \right) = \mathcal{L}\left(T_{ x^\prime}\phi^{\prime A}, x^\prime \right) + \partial_b W^b\left(T_{ x^\prime}\phi^{\prime A}, x^\prime \right) \,,
\end{equation}
where $\,W^b\left(T_{ x^\prime}\phi^{\prime A},x^\prime \right)\,$ is a first order quantity fulfilling the asymptotic condition (\ref{TD3}).  Recall that being a Noether symmetry is a sufficient (but not necessary) condition for a transformation to preserve the field equations.

\vspace{0.1cm}
As $\,S^\prime (\mathcal{V}^\prime) = S(\mathcal{V})\,$, we have that 
\begin{equation}  \label{A4}
 \int_{\mathcal{V}^\prime} \D  x\,\mathcal{L}^\prime\left(T_x\phi^{\prime A}, x\right) -\int_{\mathcal{V}}\D  x\,\mathcal{L}\left(T_x\phi^A, x\right)  = 0
\end{equation}
where we have replaced the dummy variable $ x^\prime$ with $ x\,$.  
As depicted in Figure 1, the volumes $\mathcal{V}$ and $\mathcal{V}^\prime$ differ slightly: they share a large common part  $\mathcal{V}_0$ and differ in an infinitesimal layer close to the boundary $\partial \mathcal{V}$.
\begin{figure}  \label{f1}
\begin{center}
\includegraphics[width=12cm]{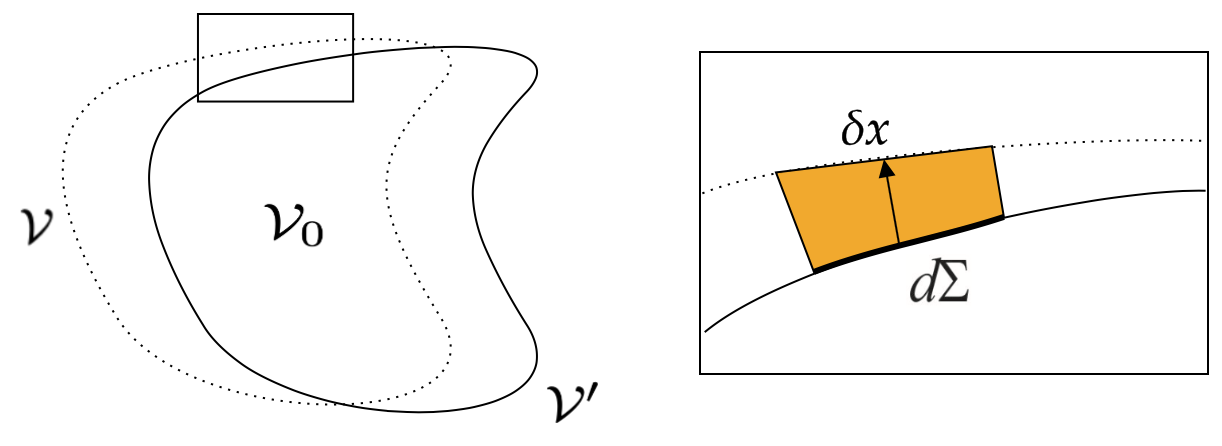} 
\end{center}
\caption{The variation of the spacetime domain $\mathcal{V}$}
\end{figure}
If $\,\D \Sigma_a\,$ is the hypersurface element on the boundary, then the volume element close to the boundary is $\; \D  x = \D \Sigma_b\,\delta  x^b \,$. Hence, by neglecting second order infinitesimals, equation (\ref{A4}) becomes
\begin{equation}  \label{A5}
 \int_{\mathcal{V}} \D x\,\left[ \mathcal{L}^\prime\left(T_x\phi^{\prime A}, x\right) - \mathcal{L}\left(T_x\phi^A, x\right)\right]\, + \int_{\partial \mathcal{V}} \mathcal{L}(T_x\phi^A, x)\,\delta  x^b \,\D \Sigma_b = 0\,.
\end{equation}
For a Noether symmetry, we have that
\begin{eqnarray*}
 \mathcal{L}^\prime\left(T_x\phi^{\prime A}, x\right) - \mathcal{L}\left(T_x\phi^A, x\right)   
 & = &  \mathcal{L}\left(T_x\phi^{\prime A}, x\right) - \mathcal{L} \left(T_x\phi^A, x\right) + \partial_b W^b\left(T_x\phi^{\prime A}, x\right) \\[1ex]
 & = & \partial_b W^b\left(T_{ x}\phi^A, x\right) + \int_{\mathbb{R}^4} \D  y\,\lambda_A(\phi, x, y)\,\delta \phi^A( y)  \,,
\end{eqnarray*}
where $\,\lambda_A(\phi, x, y)\,$ is defined in (\ref{L2o}), $\partial_b$ is the partial derivative with respect to $x^b$, and second-order terms have been neglected.
 
\vspace{0.1cm}
Introducing the variable $z= y - x\,$ in the latter, substituting it in (\ref{A5}), and applying the Gauss theorem, we obtain that
$$ \int_{\mathcal{V}}\D  x\, \left\{ \partial_b\left[\mathcal{L}(T_x\phi^A, x)\,\delta x^b + 
 W^b\left(T_x\phi^A, x\right) \right] + \int_{\RR^4} \D z\,\lambda_A(\phi, x,z+ x)\,\delta\phi^A(z+ x) \right\} = 0 $$
and, including (\ref{L2o}), we can write
\begin{eqnarray} 
\lefteqn{ - \int_{\mathcal{V}}\D  x\,\psi_A(\phi, x) \,\delta\phi^A( x) = \int_{\mathcal{V}}\D  x\, \left\{\partial_b\left(\mathcal{L}(T_x\phi^A, x)\,\delta  x^b\right) + \right.} \nonumber \\[2ex]   \label{A6}
 & & \left. \hspace*{4em}\int_{\mathbb{R}^4} \D z \,\left[ \lambda_A(\phi, x,z+ x)\,\delta\phi^A(z+ x) - \lambda_A(\phi, x-z, x)\,\delta\phi^A( x)   \right]\right\}\;.
\end{eqnarray}
Then, we use the identity
\begin{eqnarray*}
\lefteqn{\lambda_A(\phi, x,z+ x)\,\delta\phi^A(z+ x) - \lambda_A(\phi, x-z, x)\,\delta\phi^A( x) =} \\[1ex]
 &  =& \int_0^1 \D s \,\frac{\D\;}{\D s}\left\{\lambda_A(\phi, x+[s-1] z, x+s z)\,\delta\phi^A( x+s z) \right\}
\\[1ex]
 & =& \int_0^1 \D s \, z^b \,\frac{\partial\quad}{\partial  x^b} \left\{\lambda_A(\phi, x+[s-1]z, x+s z)\,\delta\phi^A( x+s z) \right]\} 
\end{eqnarray*}
that, combined with (\ref{A6}), leads to
\begin{equation}  \label{A7}
  \int_{\mathcal{V}}\D  x\,\left\{\psi_A(\phi, x) \,\delta\phi^A( x) +\frac{\partial\quad}{\partial  x^b} \left[\mathcal{L}\,\delta x^b + W^b + \Pi^b(\phi, x) \right] \right\} = 0\,,
\end{equation}
where $\;\mathcal{L}\,$ and $\, W^b\,$ are shorthands for $\,\mathcal{L}(T_x\phi, x)\,$ and $\,W^b(T_x\phi, x)\,$, and 
\begin{equation}  \label{A8}
\Pi^b(\phi, x) :=  \int_{\RR^4} \D z\, z^b \int_0^1 \D s\,\lambda_A(\phi, x+[s-1] z, x+s z)\,\delta\phi^A( x+s  z)\;.
\end{equation}

Now, as equation (\ref{A7}) holds  for any spacetime volume $\mathcal{V}$, it follows that
\begin{equation}  \label{A9}
N(\phi, x) := \partial_b J^b(\phi, x) +\psi_A(\phi, x) \,\delta\phi^A( x) \equiv 0  \,,
\end{equation}
where
\begin{equation}  \label{A10}
\quad J^b(\phi, x) := \mathcal{L}\,\delta x^b + W^b + 
 \int_{\RR^4} \D z\,z^b \int_0^1 \D s\,\lambda_A(\phi,x+[s-1] z, x+s z)\,\delta\phi^A(x+s z)\;. 
\end{equation}
Equation (\ref{A9}) is the  Noether identity and holds off shell, i. e. for any kinematic field $\phi$. On shell ---only for dynamic fields---, this identity implies that the current $\,J^b(\phi, x)\,$ is locally conserved
\begin{equation}  \label{A10a}
  \partial_b J^b = 0 \,.
\end{equation}

\subsection{Non-local Lagrangian densities that explicitely depend on $x^b$ } 
The locally conserved current (\ref{A10}) corresponds to field trajectories starting at $(\phi,0)\in \mathcal{K}^\prime\,$. In the case of explicit dependence on $x$ it is convenient to abide also trajectories initiating at any $\,(\phi,x)\,$. Therefore, we follow the same procedure as in Section \ref{S2} through the translation (\ref{L1a}) and the correspondence (\ref{L2z}). In this way, we obtain the extended current
$$\hat{J}^b(\phi, x,y) := J^b(T_{-x} \phi,x+y) \,,$$
that is 
\begin{eqnarray}  
\hat{J}^b(\phi, x,y) &=& \mathcal{L}(T_{y}\phi,x+y)\,\delta\left(x^b +y^b\right) + W^b(T_{y}\phi,x+y) + \nonumber \\[1ex] \label{A11}
  & & \int_{\RR^4} \D z\,z^b \int_0^1\D s\,\Lambda_A(\phi,x, y+[s-1] z, y+s z)\,\delta\phi^A(y+s z)  
\end{eqnarray}
and, from the way the current has been extended, it follows that
\begin{equation}  \label{A12}
\hat{J}^b(T_z\phi, x+z,y) = J^b(T_{-x}\phi,y+x+z) = \hat{J}^b(\phi, x,y+z) \;. 
\end{equation}

\subsection{Finite dimensional Lie groups. First Noether's theorem}
In case that the transformation (\ref{A3}) belongs to an $N$-parameter Lie group, then
$$ \delta x^b(x) = \varepsilon^\alpha \,\xi_\alpha^b(x) \,,\qquad \delta\phi^A(x) = \varepsilon^\alpha\,\Phi_\alpha^A(\phi,x) \,,\qquad W^b(x) = \varepsilon^\alpha \,W_\alpha^b(x) \,, $$
where $\,\xi_\alpha^b(x)\,$ is the infinitesimal generator for the parameter $\varepsilon^\alpha\,,\;\,\alpha = 1 \ldots N\,$. The current $J^b(x)$ can be written as
$$ J^b(x) = \varepsilon^\alpha J_\alpha^b(x) \,,\qquad {\rm with} \qquad \partial_b J_\alpha^b(x) = 0 \,, $$
and we have one conserved current for each group parameter, namely
\begin{equation}  \label{A10b}
 \quad J_\alpha^b(x) := \mathcal{L}\,\xi_\alpha^b(x) + W_\alpha^b +  \int_{\RR^4} \D z\,z^b \int_0^1 \D s\,\lambda_A(\phi,x+[s-1] z, x+s z)\,\Phi_\alpha^A(x+s z)  \,.
\end{equation}
Recall that $\,\mathcal{L} = \mathcal{L}(T_x\phi, x)\,$ and $\,W_\alpha^b = W_\alpha^b(T_x\phi, x) \,$.

\subsection{Poincar\'e invariance. Energy-momentum and angular momentum currents  \label{Poinc}}
Infinitesimal Poincar\'e  transformations act on coordinates as 
\begin{equation}  \label{P1}
x^{\prime a} = x^a + \delta x^a\,, \qquad \delta x^a = \varepsilon^a + \omega^a_{\;b} x^b\,, \qquad \omega_{ab}+\omega_{ba}=0 \,,
\end{equation}
where $\,\varepsilon^a\,$ and $\,\omega^a_{\;b}\,$ are constants, $\; \omega_{ab} = \eta_{ac}\omega^c_{\;b} \,$ and $\eta_{ac}=\mathrm{diag}(1,1,1,-1)$ is the Minkowski matrix to raise and lower indices. In turn, the field $\phi^A$ transforms as a tensor object ($A=1 \ldots n$ are the different components of the field) 
\begin{equation}  \label{P2}
\phi^{\prime\,A}(x^\prime) = \phi^A(x) + M^A_{\; B} \phi^B(x)\,, \qquad 
\end{equation}
where the constant matrix $\; M^A_{\; B} = \omega^{ab}\,M^A_{\; B[ab]} \;$ depends on the tensor type of the field. Hence,
\begin{equation}  \label{P3}
\delta \phi^A(x) := \phi^{\prime\,A}(x) - \phi^A(x) = \omega^{ab}\,M^A_{\; B[ab]} \phi^B(x) - \phi^A_{|c}(x) \,\left(\varepsilon^c + \omega^c_{\;b} x^b\right) 
\end{equation}
where  (\ref{A3}) and (\ref{P1}) have been included.

\vspace{0.1cm}
Then, substituting (\ref{P1}) and (\ref{P3}) into (\ref{A10}), and assuming that the Lagrangian density is Poincaré invariant --- therefore, $W^b = 0$---, we find that the conserved current  can be written as
\begin{equation}  \label{P4}
 J^b(\phi,x) = -\varepsilon^a\,\mathcal{T}^{\; b}_a(\phi,x) - \frac12\, \omega^{ac} \mathcal{J}^{\; \,\;b}_{ac}(\phi,x) \,,
\end{equation}
where
\begin{eqnarray}  \label{P5}
 \mathcal{T}^{\; b}_a &:=& -\mathcal{L}(T_y\phi, y)\,\delta^b_a +  
\int_{\RR^4} \D z\,z^b \int_0^1 \D s\,\lambda_A(\phi,y+[s-1]z, y+s z)\,\phi^A_{|a}(y+s z)  \,, \\[1ex] \label{P6}
  \mathcal{J}^{\; \,\;b}_{ac} &:=&  2\, y_{[c}\mathcal{T}^{\; b}_{a]} + \mathcal{S}^{\;\,\;b}_{ac}  \quad \qquad {\rm and}
	\end{eqnarray}
\begin{equation} \label{P6a}  \mathcal{S}^{\;\,\;b}_{ac} :=2 \int_{\RR^4} \D z\, z^b \int_0^1 \D s\,\lambda_A(\phi, y+[s-1]z, y+s z)\, \left[ s\,z_{[c} \phi^A_{|a]}(y+s z) - M^A_{\; B[ac]}\phi^B(y+s z) \right]
\end{equation}
are the currents of energy-momentum, angular momentum and spin, respectively.

\vspace{0.1cm}
As the ten parameters $\varepsilon^a$ and $\omega^{ac}$ are independent, the local conservation of the current $\,J^b\,$ implies that the currents $\mathcal{T}^{\; b}_a$ and $\,\mathcal{J}^{\; \,\;b}_{ac}\,$ are separately conserved, that is
$$ \frac{\partial \quad}{\partial y^b}\,\mathcal{T}^{\; b}_a(\phi,y) = 0 \qquad {\rm and}\qquad \frac{\partial \quad}{\partial y^b}\,\mathcal{J}^{\; \,\;b}_{ac}(\phi,y) = 0\,, $$
or 
\begin{equation}  \label{P7}
 \partial_b \mathcal{T}^{\; b}_a = 0 \qquad {\rm and}\qquad \partial_b\mathcal{S}^{\; \,\;b}_{ac} + 2 \,\mathcal{T}_{[ac]} = 0\,.
\end{equation}
$\mathcal{T}^{a b}\,$ is also known as the canonical energy-momentum tensor which, as a rule, is non-symmetric. As a consequence of the second equation (\ref{P7}), it is symmetric if, and only if, the divergence of the spin current vanishes. Indeed, this fact happens for scalar fields ruled by a local Lagrangian of the first order. For higher-order Lagrangians, there is a spin current even for them.

\vspace{0.1cm}
In all cases, an energy-momentum tensor $\Theta^{ab}$ can be found such that (in a well-defined sense) is equivalent to $\mathcal{T}^{a b}\,$ by means of the Belinfante-Rosenfeld technique \cite{BELINFANTE1940,rosenfeld1940,dixon1978}:
\begin{equation}
 \Theta^{a b} = \mathcal{T}^{a b} + \partial_c \mathcal{W}^{cba}
\end{equation}
with
\begin{equation}
\mathcal{W}^{cba} :=\frac12\,\left(\mathcal{S}^{cba} + \mathcal{S}^{cab} - \mathcal{S}^{bac} \right)\;. 
\end{equation}
In addition, Rosenfeld proves that, for finite-order Lagrangians, $\Theta^{a b}$ is actually the Hilbert energy-momentum tensor \cite{Landau1975}.

\vspace{0.1cm}
The expressions (\ref{P5}-\ref{P6a}) correspond to field trajectories initiating at $(\phi,0)\in\mathcal{D}^\prime$. For trajectories starting at any $(\phi,x)$, we should  use (\ref{A11}) rather than (\ref{A10}), and for the energy-momentum tensor we obtain
\begin{eqnarray}  
\hat{\mathcal{T}}^{\; b}_a(\phi,x,y) &:=& \mathcal{T}^{\; b}_a(T_{-x}\phi,x+y) = 
-\mathcal{L}(T_y\phi, x+y)\,\delta^b_a +  \nonumber \\[1ex]  \label{P8}
 & & \int_{\RR^4} \D z\,z^b \int_0^1 \D s\,\Lambda_A(\phi,x,y+[s-1]z, y+s z)\,\phi^A_{|a}(y+s z)   \\[1ex] 
\hat{\mathcal{S}}^{\;\,\;b}_{ac}(\phi,x,y) &:=& 2 \,\int_{\RR^4} \D z\, z^b \int_0^1 \D s\,\Lambda_A(\phi,x, y+[s-1]z, y+s z)\,\times \nonumber \\[1ex]  \label{P8a}
  & & \hspace*{8em} \left[ s\,z_{[c} \phi^A_{|a]}(y+s z) - M^A_{\; B[ac]}\phi^B(y+s z) \right]\,,
\end{eqnarray}
where (\ref{L2z}) has been included.

\subsection{Energy density and the Legendre transformation \label{Noe.2}}
Searching for some insights to generalize the Legendre transformation, we now examine the expression of the energy. The component $\,\mathcal{T}^{\; 4}_4(x)\,$ of the energy-momentum tensor is the energy density and, using (\ref{P8}) and putting $y^b = (\mathbf{y},\tau)\,$, the total energy for a field trajectory starting at $(\phi,\mathbf{0},t)\,$ is
\begin{equation}  \label{P9}
  E(\phi,t,\tau) := \int_{\RR^3} \D\mathbf{y}\,\hat{\mathcal{T}}^{\; 4}_4(\phi,\mathbf{0},t,\mathbf{y},\tau) = \int_{\RR^3} \D\mathbf{y}\,\mathcal{T}^{\; 4}_4(\phi,x^a + y^a)\,.
\end{equation}

It is well-known \cite{Landau1975} that, if the field decays fast enough at spatial infinity, the continuity equation (\ref{P7}) implies that the total energy and momentum do not depend on $\tau\,$. In the particular case of the energy, this fact implies that
\begin{equation}  \label{P9a}
  E(\phi,t, \tau) =  E(\phi,t, 0) =:E(\phi,t)\;.
\end{equation}
Therefore
\begin{eqnarray} 
 E(\phi,t) &:=&  - L(\phi, t) + \int_{\RR^6} \D\mathbf{y}\, \D\mathbf{z}\int_\RR \D\zeta \int_0^1\D s\,\zeta\,\dot\phi^A( \mathbf{y}+s\,\mathbf{z},s\,\zeta)\,\times 
\nonumber \\[1ex]  \label{P10}
 & & \hspace*{2em} \Lambda_A(\phi,\mathbf{0},t,\mathbf{y}+(s-1)\,\mathbf{z},(s-1)\zeta, \mathbf{y}+s\,\mathbf{z},s\,\zeta) 
\end{eqnarray}
where $\;L(\phi, t) :=\int_{\RR^3}\D\mathbf{y}\,\mathcal{L}(T_\mathbf{y}\phi,\mathbf{y},t)\; $, $\;y^a=(\mathbf{y},\tau)\;$, $z^a =(\mathbf{z},\zeta)\;$, and $\dot\phi^A = \phi^A_{|4}\,$.

\vspace{0.1cm}
After transforming the variables $\; \mathbf{u} = \mathbf{y} + s\,\mathbf{z} \;$ and $\; \rho =s\,\zeta \,$, the integral on the right-hand side becomes
$$ \int_{\RR^6} \D\mathbf{u}\, \D\mathbf{z}\int_\RR \D\zeta \int_0^{\zeta} \D\rho\,\Lambda_A(\phi,\mathbf{0},t,\mathbf{u}-\mathbf{z},\rho-\zeta, \mathbf{u},\rho)\,\dot\phi^A( \mathbf{u},\rho) = \hspace*{8em}$$
$$ \int_{\RR^4} \D u\, \dot\phi^A(u)\,\int_{\RR^4} \D z\, \left[\theta(u^4)-\theta(u^4-z^4)\right]\,
\Lambda_A(\phi,\mathbf{0},t,u-z,\zeta^\prime, u)\,,$$
where we have taken $\,u = (\mathbf{u},\rho)\,$ and $\,z = (\mathbf{z},\zeta)\,$. 

\vspace{0.1cm}
Then, going back to (\ref{P10}) and renaming the dummy variable $\zeta^\prime$ as  $\zeta$ and $\,(\mathbf{u},\rho) = y^a\,$, we arrive at
\begin{equation}  \label{P11}
  E(\phi, t)  = - L(\phi, t) + \int_{\RR^4} \D u\, \dot\phi^A(u)\,P_A(\phi,t,u) 
\end{equation}
where
\begin{equation}  \label{P11a}
P_A(\phi,t,u) := \int_{\RR^4} \D z\, \left[\theta(u^4)-\theta(u^4-z^4)\right]\,\Lambda_A(\phi,\mathbf{0},t,u-z,u) 
\end{equation}
is the momentum. 

\vspace{0.1cm}
The total energy can be written as $\;E(\phi, t) =\int_{\RR^3} \D\mathbf{x}\, \mathcal{E}(\phi,\mathbf{x}, t)\,$, where the energy density is
\begin{equation}  \label{P12}
\mathcal{E}(\phi,\mathbf{x}, t) := - \mathcal{L}(T_\mathbf{x}\phi,\mathbf{x}, t) + \int_\RR \D\rho\,\dot\phi^A( \mathbf{x},\rho) \,P_A(\phi,t,\mathbf{x},\rho)\;,
\end{equation}
and $\; \dot\phi^A( \mathbf{x},\rho) = \partial_\rho\phi^A( \mathbf{x},\rho) \;$ is understood.

\section{Hamiltonian formalism  \label{Ham}} 
We will now set up a Hamiltonian formalism for the Lagrange equations (\ref{L3}). The procedure is similar to the one designed in ref. \cite{Heredia2} for non-local Lagrangian mechanics with a finite number of degrees of freedom.

\subsection{Generalised Legendre transformation \label{Ham.1} }
We start by introducing the extended phase space $\,\Gamma^\prime =\mathcal{K}^2\times\RR\,$ consisting of the elements $(\phi, \pi, t)\,$, together with the Hamiltonian 
\begin{equation}  \label{H1} 
 H(\phi, \pi,t)= \int_{\RR^4} \D y\,\pi_A(y)\,\dot\phi^A(y) -  L(\phi,t)  \,,
\end{equation}
where $\phi=(\phi^1 \ldots\phi^m),\; \pi=(\pi_1 \ldots\pi_m) \in\mathcal{K}\,$ are smooth functions, $\,\mathcal{K} = \mathcal{C}^\infty(\RR^4,\RR^m)\,$,
and the Poisson bracket
$$ \left\{ F, G \right\} = \int_{\RR^4} \D y\,\left(\frac{\delta F}{\delta\phi^A(y)}\,\frac{\delta G}{\delta\pi_A(y)} - \frac{\delta F}{\delta \pi_A(y)}\,\frac{\delta G}{\delta\phi^A(y)} \right) \,.$$
Thus, the Hamilton equations are
\begin{eqnarray}  \label{H2a}
\mathbf{H} {\phi^A}(y) & = & \frac{\delta H}{\delta \pi_A(y)} = \dot\phi^A(y)  \\[1ex]  \label{H2b}
\mathbf{H} {\pi_A}(y) & = & -\frac{\delta H}{\delta\phi^A(y)} = \dot\pi_A(y) + \int_{\RR^3} \D\mathbf{x} \,\Lambda_A(\phi,\mathbf{0},t,\mathbf{x},0,y) \,,
\end{eqnarray}
where $\mathbf{H}$ is the generator of the Hamiltonian flow 
$$ (\phi^A, \pi_B, t)\rightarrow \left(T_\tau\phi^A,T_\tau\pi_B, t+\tau \right)\,, \qquad \quad \tau^b = \tau\,\delta_4^b \,.$$

Hamilton's equations can be written in a compact form by means of the contact 2-form
\begin{equation}  \label{H4} 
 \Omega^\prime = \Omega - \delta H \wedge \delta t\,, \qquad {\rm where} \qquad \Omega = \int_{\RR^4} \D y\;\delta \pi_A(y) \wedge \delta \phi^A(y) 
\end{equation}
is the symplectic form \cite{Marsden,Schmid}. Note that we have written``$\delta$'' for the differential on the manifold $\Gamma^\prime$ to distinguish it from the ``$\D$'' used in the notation for integrals we have adopted here. Then Hamilton's equations (\ref{H2a}-\ref{H2b}) become
\begin{equation}  \label{H5} 
\ri_{\mathbf{H}} \Omega^\prime = 0\;.
\end{equation}

So far, this Hamiltonian system in the extended phase space $\Gamma^\prime$ has little to do with the Lagrangian system (\ref{L3}) or the generator of time evolution $\mathbf{X}_4$ in the space $\mathcal{D}^\prime\,$. However, they can be connected by the injection map
\begin{equation}  \label{H6} 
 (\phi, t)\in \mathcal{D}^\prime \stackrel{j}{\longrightarrow} (\phi,\pi,t)\in\Gamma^\prime \,, \qquad  \mbox{where } \qquad \pi_A(y) := P_A(\phi,t)_{(y)} \in\mathcal{K}  \,,
\end{equation}
and $\, P_A(\phi,t)_{(y)} \,$ is the prefactor of $\dot\phi^A(y)$ in the energy (\ref{P11}), which is given by (\ref{P11a}),
\begin{equation}  \label{H7} 
 P_A(\phi,t)_{(y)} := P_A(\phi,t,y) = \int_{\RR^4} \D z\, \left[\theta(y^4)-\theta(y^4-z^4)\right]\,\Lambda_A(\phi,\mathbf{0},t,y-z,y) 
\end{equation}
where $\,y^b=(\mathbf{y},\rho)\,$.

\vspace{0.1cm}
$j$ defines a 1-to-1 map from $\mathcal{D}^\prime$ into its range, $j(\mathcal{D}^\prime)\subset \Gamma^\prime$, i. e. the submanifold implicitly defined by the constraints
\begin{equation}  \label{H8} 
\Psi_A\left(\phi, t\right)= 0  \qquad {\rm and} \qquad 
\Upsilon_A\left(\phi, \pi, t\right):= \pi_A - P_A(\phi,t) = 0\;.
\end{equation}
\begin{proposition}     
The Jacobian map $\,j^T\,$ maps the  infinitesimal generator $\mathbf{X}$ of time evolution in $\mathcal{D}^\prime$ into $\mathbf{H}$, i. e. the generator of the Hamiltonian flow in $\,\Gamma^\prime\,$. 
\end{proposition}

\paragraph{Proof:} To begin with, including (\ref{H6}) and (\ref{L5}), we have that
$$\; \left(j^T\mathbf{X}_4\right) \phi^A (y)= \mathbf{X}_4 \phi^A(y) = \partial_\tau\phi^A(y) =\mathbf{H}\phi^A(y) $$
where we have taken $\,y^b=(\mathbf{y},\rho)\,$. Then 
$$\left(j^T\mathbf{X}_4\right) \pi_A(y) = \mathbf{X}_4 P_A(\phi,t,y) = \left[\partial_\varepsilon P_A(T_\varepsilon\phi,t+\varepsilon,y)\right]_{\varepsilon=0} $$
and, using (\ref{H7}), (\ref{L2y}), and (\ref{H2b}), we obtain 
\begin{eqnarray*}
 \left(j^T\mathbf{X}_4\right) \pi_A(y) &=& \left[\partial_\varepsilon \int_{\RR^3} \D\mathbf{z}\int_\RR \D\zeta \,\left[\theta(\rho)-\theta(\rho-\zeta)\right]\,\Lambda_A(T_\varepsilon\phi,\mathbf{0},t+\varepsilon,\mathbf{y}-\mathbf{z},\rho-\zeta,\mathbf{y},\rho)\right]_{\varepsilon=0} \\[1ex]
 &=& \left[\partial_\varepsilon \int_{\RR^3} \D\mathbf{z}\int_\RR \D\zeta \,\left[\theta(\rho)-\theta(\rho-\zeta)\right]\,\Lambda_A(\phi,\mathbf{0},t,\mathbf{y}-\mathbf{z},\rho-\zeta+\varepsilon,\mathbf{y},\rho+\varepsilon)\right]_{\varepsilon=0} \\[1ex]
 &=&  \int_{\RR^3} \D\mathbf{z}\int_\RR \D\zeta^\prime \,\partial_\varepsilon \left(\left[\theta(\rho)-\theta(\zeta^\prime-\varepsilon)\right]\,\Lambda_A(\phi,\mathbf{0},t,\mathbf{y}-\mathbf{z},\zeta^\prime,\mathbf{y},\rho+\varepsilon)\right)_{\varepsilon=0} \\[1ex]
&=&  \partial_\rho P_A(\phi,t,\mathbf{y},\rho) + \int_{\RR^3} \D\mathbf{x}\,\Lambda_A(\phi,\mathbf{0},t,\mathbf{x},0,\mathbf{y},\rho)  \\[1ex]
 &=& \mathbf{H} \pi_A(\mathbf{y},\rho) 
\end{eqnarray*}
where we have successively taken  $\,y^b=(\mathbf{y},\rho)\,$, $\;\mathbf{x}=\mathbf{y}-\mathbf{z}\,$, $\,\zeta^\prime=\rho-\zeta-\varepsilon\,$, and have used that the second term in the right-hand side of the last but one line vanishes because the generator $\mathbf{X}_4\,$ is a solution of the field equations (\ref{L3}). \hfill $\Box$

\vspace{0.1cm}
As a corollary, $\mathbf{H}= j^T\mathbf{X}_4$ is tangent to the submanifold $j(\mathcal{D}^\prime)$, and therefore the constraints (\ref{H6}) are stable by the Hamiltonian flow.

\vspace{0.1cm}
To translate the Hamiltonian formalism in $\Gamma^\prime$ into a Hamiltonian formalism in the extended dynamic space $\mathcal{D}^\prime$, we use that the pullback $j^\ast$ maps the contact form (\ref{H4}) onto the differential 2-form 
\begin{equation} \label{H9}
\omega^\prime  = j^\ast \Omega^\prime = \int_{\RR^4} \D y\,\delta P_A(\phi,t,y)\wedge \delta \phi^A(y) - \delta h \wedge \delta t \,, \qquad \quad \omega^\prime\in\Lambda^2(\mathcal{D}^\prime)\,,
\end{equation} 
where $\,h = H\circ j\,$. Then, since $\,j^T\mathbf{X}_4 = \mathbf{H}\,$, the pullback of equation (\ref{H5}) implies that  
\begin{equation} \label{H11}
\ri_{\mathbf{X}_4} \omega^\prime = 0\;. 
\end{equation}

The reduced Hamiltonian $h(\phi,t)$ and the contact form $\omega^\prime$ on $\mathcal{D}^\prime\,$ are derived using equations (\ref{H1}) and (\ref{H7}), and they are 
\begin{equation}  \label{H12} 
h(\phi,t) =  - L(\phi,t) + \int_{\RR^8} \D y\,\D z\,\left[\theta(y^4)-\theta(y^4-z^4)\right]\,\dot\phi^A(y)\,\Lambda_A(\phi,\mathbf{0},t, y-z,y)  \,, 
\end{equation}
and $\;\, \omega^\prime_{(\phi,t)} = - \delta h(\phi,t) \wedge \delta t + \omega_{(\phi,t)} \,$, where
\begin{equation}  \label{H13} 
\omega_{(\phi,t)} = \int_{\RR^8} \D y\,\D z\,\left[\theta(y^4)-\theta(y^4-z^4)\right]\,\delta\Lambda_A(\phi,\mathbf{0},t, y-z,y) \wedge \delta\phi^A(y) \,,
\end{equation}
is the (pre)symplectic form.

\vspace{0.1cm}
We have not reached our goal yet. Because the constraints that characterize the dynamic space as a submanifold of the kinematic space $\mathcal{K}^\prime\,$ are $\,\Psi_A(\phi,t)=0\,$, $\phi^A$ and $t$ are not independent coordinates in $\mathcal{D}^\prime$\,. Therefore, the remaining final step consists of coordinatizing $\mathcal{D}^\prime$. We need to obtain the explicit parametric form of the submanifold $\mathcal{D}^\prime$ instead of the implicit form provided by the Lagrange equations. This fact is easy for regular local Lagrangians that depend on derivatives up to the $n$-th order since, as Lagrange's equations are a partial differential system of order $2n$, the Cauchy-Kowalewski theorem \cite{john1991} provides the sought parametric form.
However, as a rule, deriving the explicit equations of $\mathcal{D}^\prime$ from the implicit equations is a complex task that depends on each specific case.

\vspace{0.3cm}
\noindent
Let us exemplify this fact in the next Section.

\section{Application. The $p$-adic open string field \label{S5}}
The ``user manual'' for the procedure developed so far would read:
\begin{itemize}
\item To start with, write the action integral so that the function $ \mathcal{L}(T_y \phi,y)$ can be identified, 
\item Compute the functional derivatives $\,\lambda(\phi,y,z)\,$ and $\Lambda(\phi,x,y,z)$ ---see equations (\ref{L2o}) and (\ref{L2z})---,
\item Substitute the latter in (\ref{H13}) and calculate both the contact form and 
\item The Hamiltonian (\ref{H12}).
\end{itemize}
In what follows, we shall apply these directions to the $p$-adic open string case. The main difficulty stems from finding a complete set of coordinates to characterize the elements of the dynamic space $\mathcal{D}^\prime\,$. 

\vspace{0.1cm}
\noindent
On the other hand, let us also mention that we are outsiders of this model. Our intention is only to illustrate the procedure of how to apply this formalism and not to analyze the result obtained. We leave the latter to the reader who is specialized in this field.  

\vspace{0.1cm}
We consider the Lagrangian density for the $p$-adic open string 
\begin{equation}  
\label{padic:Lnt}
\mathcal{L}(\psi) = - \frac{1}{2} \psi \:e^{-r\Box} \psi + \frac{1}{p+1} \psi^{p+1} \qquad \mathrm{with} \qquad r = \frac{1}{2} \ln(p)
\end{equation}
where $\Box$ is the d'Alembert operator $\Box = \eta^{\alpha\beta}\partial_\alpha\partial_\beta$, $\eta^{\alpha\beta}$ is the inverse Minkowski metric, and $p$ is a prime number. 

\vspace{0.1cm}
As for the kinematic space, $\psi(y)$ is a smooth function $\mathcal{C}^\infty(\mathbb{R}^{3+1})$ such that the operation 
$$e^{-r\Box} \psi(y) := \sum_{n=0}^\infty \frac{(-r)^n}{n!}\,\Box^n\psi(y)$$ 
is ``well-defined''. Since it is a part of the Lagrangian density (\ref{padic:Lnt}), we need the series in $e^{-r\Box} \psi(y)$ to converge in some appropriate functional space. The consequences of this requirement have been thoroughly analysed in ref. \cite{heredia2021} and led to the conclusion that:

\begin{list}
{(\alph{llista})}{\usecounter{llista}} 
\item Each function $\psi(x)$ in the kinematic space is the result of a convolution
\begin{equation}  \label{p2}
\psi(x) := (\E \ast\:\phi)(x) \,, \qquad {\rm where} \qquad \E(\mathbf{x},t):= \G_3(\mathbf{x})\,\delta(t)
\end{equation}
and $x:= (\mathbf{x},t)$, for some smooth function $\phi(\mathbf{x},t)$ that grows slowly at $|\mathbf{x}|\rightarrow\infty\,$, that is 
$\;\forall \alpha = (\alpha_1, \ldots \alpha_n)\,, \; \alpha_j = 1 \ldots 4\,$ , they exist 
$$ C_\alpha(t)> 0\quad {\rm and} \quad m_\alpha(t) \in \mathbb{Z}^+ \quad \mbox{such that} \quad \left|\partial_{\alpha_a \ldots \alpha_n} \phi(\mathbf{x},t)  \right|\leq C_\alpha(t) \,\left(1+|\mathbf{x}|^2 \right)^{m_\alpha(t)} \,.$$
No restriction on the behavior of $\phi(\mathbf{x},t)$ at large $|t|$ is imposed. We shall denote by $\theta_{M}(\mathbb{R}^3)$ the class of these functions \cite{Vladimirov_GF}.
\item The operator $e^{-r\Box}$ acts as 
\begin{equation}  \label{p3}  
e^{-r\Box}\psi(x) := \left(\T \ast\:\phi \right)(x) \,, \qquad {\rm where} \qquad \T(\mathbf{x},t):= \delta(\mathbf{x})\,\G_1(t)\,.
\end{equation}
The functions $\G_1$ and $\G_3$ are defined by\footnote{In fact, they are the heat kernels in 1-dimension and 3-dimension space where $r$ plays the role of evolution parameter of the heat equation. See for instance \cite{Kolar2022}. }
\begin{equation}  \label{p5}
\G_n(x) = \frac{1}{(2\sqrt{\pi r})^n} e^{-\frac{|x|^2}{4r}}, \qquad x \in \RR^n \,, \qquad n= 1\;{\rm or}\; 3\,.
\end{equation}
\end{list}
Including all this, in terms of the new kinematic variables $\phi(y)\,$, the Lagrangian (\ref{padic:Lnt}) becomes
\begin{equation}
\label{padic:Lntc}
\mathcal{L}(T_y \phi,y) = - \frac{1}{2}\left(\E \ast\:\phi\right)(y) \: \left(\T \ast\:\phi\right)(y) + \frac{1}{p+1}\left[(\E \ast\: \phi)(y)\right]^{p+1}\;.
\end{equation}

For the Lagrangian density (\ref{padic:Lntc}), the functional derivative (\ref{L2o}) is
\begin{equation}
\label{padic:Lambda}
\begin{split}
\lambda(\phi,y,z) =& \E(y-z) \left\{ -\frac{1}{2}(\T \ast\:\phi)(y) +\left[(\E \ast\:\phi)(y)\right]^p\right\}- \frac{1}{2}  \T(y-z)\left(\E \ast\:\phi\right)(y)
\end{split}
\end{equation}
where $z:= (\mathbf{z},\xi)$, $y:= (\mathbf{y},\tau)$ and, as the Lagrangian density does not explicitly depend on $x^a$, the functional derivative (\ref{L2z}) is
$$\Lambda(\phi,x,y,z) = \lambda(\phi,y,z)\,.$$

The Euler-Lagrange equation (\ref{L2}-\ref{L2z}) easily follows and has the form of the convolution equation
\begin{equation}   \label{p6a}
\E \ast\left[\T \ast\:\phi - \left(\E \ast\:\phi\right)^p\right] = 0 \,,
\end{equation}
which amounts to ---see Appendix A.1 for details---
\begin{equation}   \label{p6}
\T \ast\:\phi - \left(\E \ast\:\phi\right)^p = 0 \,.
\end{equation}

If we now consider the spatially homogeneous case, $\phi(x) := \phi(t)$, the field equations are the same as for the $p$-adic particle case \cite{Vladimirov2004}.  A possible solution of (\ref{p6}) is
\begin{equation}
\label{padic:phisolution}
\phi_0(x) =
\left\{
	\begin{array}{ll}
		\pm 1,0  & \quad \mbox{if } p \text{ is odd} \\
		1,0 & \quad \mbox{if } p \text{ is even} \:.
	\end{array}
\right.
\end{equation}
The field equation (\ref{p6}) admits other solutions \cite{Vladimirov2012}; however, we will focus on the perturbative ones around $\phi_0(x)\neq 0 $, namely
\begin{equation}
\phi(x) = \phi_0(x) + \kappa\:\Phi(x)\;,
\end{equation}
where $\kappa\ll1$ is the expansion parameter. For the sake of simplicity, we choose $p=2$, which is even, and therefore $\phi_0(x) = 1$. Thus, substituting $\Phi(x) = \sum^\infty_{n=0} \kappa^n \Phi_n (x)$ in the field equation (\ref{p6}), we get
\begin{equation}
\label{padic:relsol}
\T \ast\:\Phi_n  - 2\,\E \ast\:\Phi_n  = \sum_{l+m = n-1} \left(\E \ast\:\Phi_l\right)\left(\E \ast\:\Phi_m\right)\:.
\end{equation}

At the lowest order, $n=0$, it reads
\begin{equation}
\label{padic:relsol0}
\T \ast\:\Phi_0  - 2\,\E \ast\:\Phi_0  = 0\:.
\end{equation}
The latter is an integral equation that might be solved using the Fourier transform but this would restrict the search to summable functions that vanish at infinity, both spatial and temporal. From a physical point of view, this makes sense for the spatial dependence, however, this restriction does not seem appropriate as far as the time dependence is concerned. For this reason, we propose that the general solution is a superposition of ``monochromatic'' solutions such as $\Phi_0(z) = e^{i \alpha\xi} \tilde{A}(\mathbf{z})$, where $\tilde{A}(\mathbf{z})$ is a summable function and $\,z^a =(\mathbf{z},\xi)\,$. Therefore, using that
\begin{equation}
\label{padic:relation}
\G_{1}(\xi)\ast\:e^{i \alpha \xi} = e^{-r \alpha^2 + i \alpha \xi}
\end{equation}
and plugging $\Phi_0(z)$ into $(\ref{padic:relsol0})$, we get
\begin{equation}
\label{padic:FT}
e^{i\alpha\xi}\,\left(e^{-r\alpha^2}\,\tilde{A}(\mathbf{z}) - 2 \,(\E \ast \tilde{A} )(\mathbf{z}) \right) = 0  \:,
\end{equation}
whose spatial Fourier transform\footnote{See Appendix for the Fourier transform convention.} yields
\begin{equation} 
\label{padic:cond}
{A}(\mathbf{k})\,e^{i\alpha\xi}\,\left(e^{-r \alpha^2 } - 2\:e^{-r|\mathbf{k}|^2}\right) = 0\:.
\end{equation}

Whereas ${A}(\mathbf{k}) = 0$ leads to the trivial solution, non-trivial solutions are connected with the spectral equation
\begin{equation} \label{spect1}
e^{-r\alpha^2 } - 2\:e^{-r|\mathbf{k}|^2}= 0 \, ,
\end{equation}
whose solution is the set of complex numbers
\begin{equation} \label{spect2}
\mathcal{N}= \left\{\alpha_\nu(\mathbf{k}) = s\, \sqrt{|\mathbf{k}|^2 - 2\left(1 + \frac{i\pi l}{r}\right)} \,, \quad \mathbf{k}\in \RR^3 \,, \quad \nu=(s,l)\,, \;\, s =\pm\,,\;\, l \in \ZZ \right\} \,.
\end{equation}
We will write
$$ \nu^\prime =(-s,l)  \,, \qquad \quad \tilde\nu=(s,-l) \,, \qquad \quad -\nu=(-s,-l)\;,$$
and we have that
\begin{equation}
\label{padic:alphas}
\alpha_{\tilde\nu}(\mathbf{k}) = \overline\alpha_\nu(\mathbf{k}) \,, \qquad \alpha_{\nu^\prime}(\mathbf{k}) = -\alpha_\nu(\mathbf{k}) \,.
\end{equation}
Therefore, the general solution of (\ref{padic:relsol0}) is
\begin{equation}  \label{p10}
\Phi_0(z) = \frac{1}{(2\pi)^3} \intRRR \D\mathbf{k} \sum_{\nu} A_\nu(\mathbf{k}) e^{i [\alpha_\nu(\mathbf{k})\xi + \mathbf{k} \cdot \mathbf{z}]}
\end{equation}
and, as $\Phi_0(z)$ has to be real,
\begin{equation}
\label{padic:As}
 A_{-\nu}(-\mathbf{k}) = \overline{A}_{\nu}(\mathbf{k})\,.
\end{equation}
Notice that, as $\alpha(\mathbf{k})$ is complex, the integral might diverge at $|\mathbf{k}| \rightarrow \infty$; however, this is not the case, as shown in Appendix A.2\,.

\vspace{0.1cm}
At the next perturbative order, $n=1$, equation (\ref{padic:relsol}) yields
\begin{equation}
\label{padic:relsol1}
\T \ast\:\Phi_1 - 2\, \E\ast\:\Phi_1 =  \left(\E \ast\:\Phi_0\right)^2   \:.
\end{equation}
Using (\ref{padic:relation}), the right-hand side of this equation can be written as
\begin{equation}
\left(\E \ast\:\Phi_0\right)^2(z) = \int_{\mathbb{R}^6} \frac{\D\mathbf{k} \D\mathbf{p}}{(2\pi)^6}\,\sum_{\nu,\mu} A_\nu(\mathbf{k}) A_\mu(\mathbf{p})\,
 e^{-r \left(|\mathbf{k}|^2 + |\mathbf{p}|^2 \right)+ i (\mathbf{k} + \mathbf{p})\cdot\mathbf{z}}\, 
e^{i \left[\alpha_\nu(\mathbf{k})+\alpha_\mu (\mathbf{p})\right]\xi} \,.
\end{equation}

Again, a particular solution (\ref{padic:relsol1}) can be obtained as a superposition of ``monochromatic'' solutions like 
$\,\displaystyle{ \tilde\Phi_1(\mathbf{k},\mathbf{p})\,e^{i (\mathbf{k}+\mathbf{p})\cdot \mathbf{z}} \, e^{i \left[\alpha_\nu(\mathbf{k})+\alpha_\mu (\mathbf{p})\right]\xi }}$. Following the same steps as above, we arrive at
\begin{equation}
\Phi_1(z) = \int_{\mathbb{R}^6} \frac{\D\mathbf{k} \D\mathbf{p}}{(2\pi)^6} \, \sum_{\nu,\mu} \frac{A_\nu(\mathbf{k}) A_\mu(\mathbf{p})}{f_{\nu\mu}(\mathbf{k},\mathbf{p})} e^{i\,[\alpha_\nu(\mathbf{k})+\alpha_\mu(\mathbf{p})]\xi + i\,(\mathbf{k} +\mathbf{p})\cdot\mathbf{z}}  \,,
\end{equation}
where
\begin{equation}
f_{\nu\mu}(\mathbf{k},\mathbf{p}):= 2 \left[2e^{-2r \alpha_\nu(\mathbf{k})\alpha_\mu(\mathbf{p})}- e^{-2r\mathbf{k}\cdot\mathbf{p}}\right]
\end{equation}
and we have used that $\,\alpha_\nu(\mathbf{k})\,$ is a solution of the spectral equation (\ref{spect1}). Therefore, the general perturbative solution up to the second-order is
\begin{equation}  \label{pertsol}
\phi(\mathbf{z},\xi) = 1 + \kappa\:\intRRR \frac{\D \mathbf{k}}{(2\pi)^3} \sum_{\nu} A_\nu(\mathbf{k}) e^{i [\alpha_\nu(\mathbf{k})\xi + \mathbf{k} \cdot \mathbf{z}]}\,
\left\{1 + \kappa \intRRR \frac{\D \mathbf{p}}{(2\pi)^3}\sum_{\mu} \frac{A_\mu(\mathbf{p})}{f_{\nu\mu}(\mathbf{k},\mathbf{p})} e^{i\,[\alpha_\mu(\mathbf{p})\xi + \mathbf{p}\cdot\mathbf{z}]} \right\} \,.
\end{equation}

\subsection{The symplectic form \label{SS5.1}}
Using (\ref{padic:Lambda}), we find that the momentum (\ref{H7}) is
\begin{equation}
\label{padic:momenta}
P(\phi,y) = -\frac{1}{2} \intR \D\xi \left[\theta(\tau) - \theta(\xi)\right] \G_1(\xi-\tau) \left(\E \ast\:\phi \right)(\mathbf{y},\xi)
\end{equation}
and, therefore, the (pre)symplectic form (\ref{H13}) becomes
\begin{equation}
\label{padic:simpletic0}
\omega = - \frac{1}{2} \int_{\mathbb{R}^4} \D\mathbf{y}\:\D s \G_1(s) \int^s_0 \D\tau \left(\E \ast\:\delta\phi\right)(\mathbf{y},\tau-s) \wedge \delta\phi(\mathbf{y},\tau)\:,
\end{equation}
where we have introduced the change $s =\xi-\tau$. Taking now 
$$\phi(z)= 1 + \kappa \Phi_0(z) + \kappa^2 \Phi_1(z) + \mathcal{O}(\kappa^3) \,,$$ 
after a bit of algebra ---see Appendix A.3 for details---, we obtain that the symplectic form is 
\begin{equation}
\label{padic:simpletic}
\omega = i\,\intRRR \D\mathbf{k}\,\sum_{l\in\mathbb{Z}} \delta B_l(\mathbf{k})\wedge \delta B^\dagger_l(\mathbf{k}) + \mathcal{O}(\kappa^4) 
\end{equation}
where the new variables 
\begin{equation}  \label{f3p}
B_l(\mathbf{k}):= \frac{2\,\kappa\,e^{-r|\mathbf{k}|^2}}{(2\pi)^{3/2}} \,\sqrt{r\: \alpha_l(\mathbf{k})} \, A_{(+,l)}(\mathbf{k}) \qquad {\rm and} \qquad B^\dagger_l(\mathbf{k}):= \overline{B}_{-l}(\mathbf{k}) 
\end{equation}
have been introduced.

\vspace{0.1cm}
It is apparent that: (a) $\omega$ is non-degenerate, hence symplectic, and (b) the modes $B_l(\mathbf{k})$ and $ B^\dagger_{j}(\mathbf{k})$ are a  system of canonical coordinates whose elementary Poisson brackets are
\begin{equation}  \label{f4p}
 \{B_l(\mathbf{k}),B^\dagger_{j}(\mathbf{k}^\prime)\} = i\,\delta_{lj}\,\delta(\mathbf{k}-\mathbf{k}^\prime) + \mathcal{O}(\kappa^4)\,, \qquad 
\{B_l(\mathbf{k}), B_{j}(\mathbf{k}^\prime)\} = \{B^\dagger_l(\mathbf{k}), B^\dagger_{j}(\mathbf{k}^\prime)\} =  \mathcal{O}(\kappa^4) \,.
\end{equation}

\subsection{The Hamiltonian \label{SS5.2}}
Substituting  the momenta (\ref{padic:momenta}) in equation (\ref{H12}), we obtain that the Hamiltonian is
\begin{equation}   \label{ham1}
h(\phi) = - L(\phi) + \frac12\intRRR \D\mathbf{y} \intR \D s \G_1(s) \int^s_0 \D\xi\:\dot\phi(\mathbf{y},\xi-s)\left(\E \ast\:\phi\right)(\mathbf{y},\xi)
\end{equation}
where we have defined $s=\xi-\tau$ and
\begin{equation}
\label{ham0}
L(\phi) := \intRRR \D \mathbf{y}\, \mathcal{L}(T_\mathbf{y} \phi,\mathbf{y},0) 
\end{equation}
with $\mathcal{L}$ given by (\ref{padic:Lntc}).

\vspace{0.1cm}
Again, taking the perturbative expansion of $\phi(z)$ up to $\kappa^2$ terms, we obtain that the Hamiltonian is ---see Appendix A.4 for details---
\begin{equation}  \label{ham1a}
h = \frac{V_y}{6} + \intRRR  \D\mathbf{k}\,\sum_{l\in\mathbb{Z}} \alpha_l(\mathbf{k})\:B_l(\mathbf{k}) B^\dagger_l(\mathbf{k}) + \mathcal{O}(\kappa^3) \,,
\end{equation}
where $V_y =\intRRR \D\mathbf{y}$ is an infinite contribution to the Hamiltonian, which is associated with the (divergent) vacuum energy. This problem may be highly complex to treat when gravity is present \cite{Martin2012}. However, as gravity is absent in our case, we can simply drop it \cite{Peskin1995}. Moreover, since the Hamiltonian is treated as the generator of the dynamics of our system, Hamilton's equations will not be affected by this term because it is merely a constant. Therefore, the Hamilton equations for this Hamiltonian with the Poisson brackets (\ref{f4p}) are 
\begin{equation}
\mathbf{h} B_j(\mathbf{k}) = i\:\alpha_j(\mathbf{k}) \,B_j(\mathbf{k})\qquad {\rm and} \qquad \mathbf{h} B^\dagger_j(\mathbf{k}) = - i\:\alpha_j(\mathbf{k}) \,B^\dagger_j(\mathbf{k}) \:.
\end{equation}

\subsection{The energy-momentum tensor \label{SS5.3}}
In \cite{Moeller2002}, an expression of the energy-momentum tensor for the homogeneous infinite-order $p$-adic Lagrangian is obtained in a non-closed form (i.e. expressed as an infinite series).  Since our formalism allows us to calculate both the canonical energy-momentum tensor $\hat{\mathcal{T}}^{a b}$ and the Belinfante-Rosenfeld energy-momentum tensor $\Theta^{a b}$ in a closed form (i.e. with the infinite series summed), we will now particuliarise these expressions for the perturbative $p$-adic open string case, as we did for non-local dispersive media \cite{Heredia1}. 

\vspace{0.1cm}
\noindent
Before calculating these tensors, we must take into account these two observations:

\vspace{0.2cm}
a) The Lagrangian density (\ref{padic:Lnt}) is Poincaré invariant if the $\psi$  field transforms as a scalar, i.e. $\psi^\prime(x^\prime) = \psi(x)$. However, note that the $\phi$ field cannot be a Poincaré scalar because its definition (\ref{p2}) is not. Therefore, we will require that the $\psi$ field transforms as a scalar to obtain the  transformation rule of $\phi$ that leaves the Lagrangian density (\ref{padic:Lntc}) Poincaré invariant (i.e., with $W^b=0$). Indeed, this transformation is --- see Appendix A.5 for details ---
\begin{equation}
\label{padic:t1_2}
 \delta \phi(x) = -\left(\varepsilon^c+ \omega^{cb} x_b\right)\partial_c\phi(x) + \omega^{ab}\left[2r\,\delta^4_{[a} \delta^i_{b]}\, \partial_i\dot\phi(x)\right]\,.
\end{equation}
As one can observe, the last term causes the $\phi$ field not to be a Poincaré scalar. In fact, this term will contribute to the spin part. Fortunately, due to the structure of this additional term, we do need to recalculate Section 3.3 since the structure of (\ref{padic:t1_2}) is equivalent to (\ref{P3}). We just need to change the $\omega^{ab} M^A_{B[ab]}\phi^b$ term by the last term of (\ref{padic:t1_2}).  

\vspace{0.2cm}
b) Both the canonical and the Belinfante-Rosenfeld energy-momentum tensor are conserved only on shell, namely, at any solution of (\ref{p6}); therefore, we have to consider the $p$-adic Lagrangian density and $\lambda(\phi,y,z)$ on shell to obtain them, that is 
\begin{equation}
\label{Los}
 \Xi(T_y\phi,y):=\mathcal{L}_{(os)}(T_y \phi,y) = -\frac{1}{6} \left[(\E \ast\:\phi)(y)\right]^3
\end{equation}
and 
\begin{equation}
\label{Laos}
\Upsilon(y,z):= \lambda_{(os)}(\phi,y,z) =\frac{1}{2}\left[\E(y-z)(\T \ast\:\phi)(y) - \T(y-z)(\E \ast\:\phi)(y)\right]
\end{equation}
respectively. As mentioned above, since the Lagrangian density does not explicitly depend on the spacetime coordinates, the canonical energy-momentum tensor $\hat{\mathcal{T}}^{a b}$ coincides with $ \mathcal{T}^{a b}$. The same is true for the spin tensor $\hat{\mathcal{S}}^{acb}$ and $\mathcal{S}^{acb}$ . 

\vspace{0.2cm}
Thus, bearing in mind the second observation and using (\ref{Los}) and (\ref{Laos}), the canonical energy-momentum tensor in closed form is
\begin{equation}
\label{padic:tc}
\mathcal{T}_a^{\:\:b}(\phi,y) = - \Xi(T_y\phi,y)\:\delta^b_a + \int^1_0 \D s \int_{\RR^4} \D z \:\Upsilon(y+[s-1]z, y+sz)\:z^b\:\phi_{|a}(y+sz)\:.
\end{equation}
Now, using (\ref{Laos}) and bearing in mind both the first and second observation, the spin current (\ref{P6a}) is
\begin{equation}
\label{padic:Sc}
\begin{split}
\mathcal{S}^{\;\,\;b}_{ac}(\phi,y) =  2 \int_{\RR^4} \D z\,z^b\,\int_0^1 \D s\,&\Upsilon(y+[s-1]z, y+sz)\\
&\times \left[s\,z_{[c}\phi_{|a]}(y+sz) + 2\,r\,\delta^4_{[c}\delta^i_{a]}\,\dot \phi_{|i}(y+sz)\right]\,.
\end{split}
\end{equation}
With the last expression (\ref{padic:Sc}), we find that $\mathcal{W}^{cba}(\phi,y)$ is
\begin{equation}
\label{padic:W}
\begin{split}
\mathcal{W}^{cba} =& \int_{\RR^4} \D z\int_0^1 \D s\,\Upsilon(y+[s-1]z, y+sz)\left[s\,(z^a z^b\,\delta^c_g -z^az^c\,\delta^b_g)\,\phi^{|g}(y+sz)\right.\\
&\left. \qquad\quad+2\,r\left(z^{(a}\eta^{b)4}\eta^{cf} - z^{(a}\eta^{b)f}\eta^{4c} - z^c\eta^{4[a}\eta^{b]f}\right) \dot \phi_{|f}(y+sz)\right]
\end{split}
\end{equation}
where we have used the fact that $\eta^{4[c}\eta^{a]i}\,\dot \phi_{|i}(y+sz) = \eta^{4[c}\eta^{a]f}\, \dot \phi_{|f}(y+sz)$ because of the antisymmetry.  Finally, deriving with respect to $y^c$ of equation (\ref{padic:W}), and using the property $z^c\, A_{|_c}(y+s z) = \frac{\D}{\D s}A(y+sz)$ and an integration by parts, we obtain
\begin{equation}
\label{padic:cW}
\begin{split}
&\partial_cW^{cba} =- \int_{\RR^4} \D z\left\{ z^a\,\Upsilon(y, y+z)\phi^{|b}(y+z) +2\,r\,\eta^{4[a}\left( \Upsilon(y, y+z) \dot \phi^{|b]}(y+z) - \Upsilon(y-z, y) \dot \phi^{|b]}(y) \right) \right.\\
&\left. - \int_0^1 \D s\left[ s\,z^a z^b\,\partial_c[\Upsilon(y+[s-1]z, y+sz)\phi^{|c}(y+sz)] +z^a\Upsilon(y+[s-1]z, y+sz)\phi^{|b}(y+sz)\right.\right.\\
&\left.\left. + 2\,r\,z^{(a}\left(\eta^{b)4}\,\partial^f[\Upsilon(y+[s-1]z, y+sz)\dot \phi_{|f}(y+sz)] -\partial^4[\Upsilon(y+[s-1]z, y+sz)\dot \phi^{|b)}(y+sz)] \right)\right] \right\}\,.
\end{split}
\end{equation}
Therefore, putting (\ref{padic:tc}) and (\ref{padic:cW}) together, we obtain the Belinfante-Rosenfeld tensor in closed form
\begin{equation}
\begin{split}
&\Theta^{ab}(\phi,y) = - \Xi(T_y\phi,y) \delta^{ab} - \int_{\RR^4} \D z\left\{\Upsilon(y, y+z) z^a \phi^{|b}(y+z) + 2\,r\,\eta^{4[a}\left(\Upsilon(y, y+z)\,\dot \phi^{|b]}(y+z)\right.\right.\\
&\left.\left.  - \Upsilon(y-z, y)\, \dot \phi^{|b]}(y) \right) -  \int_0^1 \D s\left[ s\,z^a z^b\,\partial_c[\Upsilon(y+[s-1]z, y+sz)\phi^{|c}(y+sz)] \right.\right.\\
&\left.\left. +2\,\Upsilon(y+[s-1]z, y+sz)\,z^{(a}\phi^{|b)}(y+sz)+ 2\,r\,z^{(a}\left(\eta^{b)4}\,\partial^f[\Upsilon(y+[s-1]z, y+sz)\dot \phi_{|f}(y+sz)]\right.\right.\right.\\
&\left.\left.\left. - \partial^4[\Upsilon(y+[s-1]z, y+sz) \dot \phi^{|b)}(y+sz)] \right)\right]\right\}\,.
\end{split}
\end{equation}
It is worth mentioning that the Belinfante-Rosenfeld tensor is dependent on the solution that the theory might present. For this reason, we use the pertubative solution (\ref{pertsol}) to obtain explicitly its components. 

\vspace{0.1cm}
The first element to be calculated is the (4,4)-component, which indicates the energy density of the system associated to the perturbative solution. Therefore, the energy density is
\begin{equation}
\begin{split}
&\Theta^{\,\,\,4}_4(\phi,y) = - \Xi(T_y\phi,y) + \int_0^1\D s\int_{\RR^4} \D z\,z^4\left\{\Upsilon(y+[s-1]z, y+sz)\dot\phi(y+sz)\right.\\
&\left. - s\,z^i\,\partial_i\left[\Upsilon(y+[s-1]z, y+sz)\dot\phi(y+sz)\right] - \partial_i\left[s\,z^4\,\Upsilon(y+[s-1]z, y+sz)\phi^{|i}(y+sz)\right.\right.\\
&\left.\left. - 2\,r\, \Upsilon(y+[s-1]z, y+sz) \dot\phi^{|i}(y+sz)\right] \right\}
\end{split}
\end{equation}
and, including
\begin{equation}
\label{padic:Ups}
\Upsilon(y+[s-1]z, y+sz) = \frac{1}{2}\left[\delta(\xi)\mathcal{G}_3(\mathbf{z})(\mathcal{T}\ast\phi)(y+[s-1]z) - \delta(\mathbf{z})\mathcal{G}_1(\xi)(\mathcal{E}\ast\phi)(y+[s-1]z)\right]\,,
\end{equation}
 it simplifies as
\begin{equation}
\begin{split}
\Theta^{\,\,\,4}_4(\phi,y) =& - \Xi(T_y\phi,y) - \frac{1}{2} \int_0^1\D s \int_\RR \D\xi\,\xi\,\mathcal{G}_1(\xi)\left\{(\mathcal{E}\ast\phi)(\mathbf{y},\tau+[s-1]\,\xi)\,\dot\phi(\mathbf{y},\tau+s\,\xi)\right.\\
&\left. -\partial_i\left[(\mathcal{E}\ast\phi)(\mathbf{y},\tau+[s-1]\xi)\left(s\,\xi\,\phi^{|i}(\mathbf{y},\tau+s\,\xi) -2\,r\,\dot\phi^{|i}(\mathbf{y},\tau+s\,\xi)\right)\right] \right\}\,.
\end{split}
\end{equation}
Taking (\ref{ecp3}), (\ref{padic:convh}) and (\ref{pertsol}), and computing the integrals , it becomes --- after a tedious computation---
\begin{equation}
\Theta_{4}^{\:\:4}(y) = \frac{1}{6}+ \kappa\:\overset{1}{\Theta}\,_{4}^{\:\:4}(y) + \kappa^2\:\overset{2}{\Theta}\,_{4}^{\:\:4}(y) + \mathcal{O}(\kappa^3)
\end{equation}  
where 
\begin{equation}
\begin{split}
\overset{1}{\Theta}\,_{4}^{\:\:4}(y):= \frac{1}{2} \int_{\RR^3}\frac{\D \mathbf{k}}{(2\pi)^3}& \sum_\nu A_\nu(\mathbf{k}) \:e^{i(\alpha_\nu(\mathbf{k})\tau + \mathbf{k}\cdot\mathbf{y})}\\
&\times \left[1-e^{-r|\mathbf{k}|^2} - \frac{|\mathbf{k}|^2}{2\,\alpha_\nu(\mathbf{k})^2}\left(e^{-r|\mathbf{k}|^2}+r\,\alpha_\nu(\mathbf{k})^2 - \frac{1}{2}\right)\right]
\end{split}
\end{equation}
and
\begin{equation}
\begin{split}
\overset{2}{\Theta}\,_{4}^{\:\:4}(y)&:= \frac{1}{2} \int_{\RR^6}\frac{\D \mathbf{k}\:\D\mathbf{p}}{(2\pi)^6} \sum_{\nu,\mu}A_\nu(\mathbf{k})A_\mu(\mathbf{p})\,e^{i([\alpha_\nu(\mathbf{k})+\alpha_\mu(\mathbf{p})]\tau + [\mathbf{k}+\mathbf{p}]\cdot\mathbf{y})} \left[\frac{1-e^{-r(\mathbf{k}+\mathbf{p})^2}}{f_{\nu\mu}(\mathbf{k},\mathbf{p})} \right.\\
&\quad \left. - \frac{(\mathbf{k}+\mathbf{p})^2}{f_{\nu\mu}(\mathbf{k},\mathbf{p})[\alpha_{\nu}(\mathbf{k})+\alpha_{\mu}(\mathbf{p})]^2}\left(e^{-r[\alpha_\nu(\mathbf{k}) +\alpha_\mu(\mathbf{p})]^2} + 2r\,[\alpha_{\nu}(\mathbf{k})+\alpha_{\mu}(\mathbf{p})]^2 -1\right)\right.\\
&\quad \left. + \frac{\mathbf{k}\cdot(\mathbf{k}+\mathbf{p})\,e^{-2r|\mathbf{p}|^2}}{2\,[\alpha_{\nu}(\mathbf{k})+\alpha_{\mu}(\mathbf{p})]^2}\left(e^{-r(|\mathbf{k}|^2-|\mathbf{p}|^2)} +2r\,\alpha_\nu(\mathbf{k})[\alpha_\nu(\mathbf{k})+\alpha_\mu(\mathbf{p})] -1 \right)\right.\\
&\quad \left. + \frac{\alpha_\nu(\mathbf{k})\,e^{-r|\mathbf{p}|^2}}{[\alpha_{\nu}(\mathbf{k})+\alpha_{\mu}(\mathbf{p})]}\left(e^{-r\alpha_{\mu}(\mathbf{p})^2} - e^{-r\alpha_\nu(\mathbf{k})^2}\right) \right]\:.
\end{split}
\end{equation}
From the last two expression, it is easy to prove that, if we calculate the total energy of the system at $\tau=0$, it coincides with the Hamiltonian (\ref{ham1a}) since
\begin{equation}
\int_{\RR^3} \D \mathbf{y}\:\overset{1}{\Theta}\,_{4}^{\:\:4}(\mathbf{y},0) = 0,
\end{equation}
and 
\begin{equation}
E - \frac{V_y}{6} = h - \frac{V_y}{6}  = \int_{\RR^3} \D \mathbf{y}\:\overset{2}{\Theta}\,_{4}^{\:\:4}(\mathbf{y},0) = \int_{\RR^3}  \D\mathbf{k}\,\sum_{l\in\mathbb{Z}} \alpha_l(\mathbf{k})\:B_l(\mathbf{k}) B^\dagger_l(\mathbf{k})\:.
\end{equation}
where $V_y =\intRRR \D\mathbf{y}$ is the vacuum energy. It is important to highlight this result. Note that the total energy of the system (or the Hamiltonian for this case) is not affected by choice of tensor. As we have just shown, either through the canonical or the Belinfante-Rosenfeld energy-momentum tensor, the result remains the same, as might be expected since the total energy of the system is not modified. 

\vspace{0.1cm}
The second element is the $(i,j)$-component, which indicates the pressure of the system. Thus, the pressure is
\begin{equation}
\begin{split}
\Theta^{ij}&(\phi,y) = - \Xi(T_y\phi,y) \delta^{ij} - \int_{\RR^4} \D z\left\{\Upsilon(y, y+z) z^i \phi^{|j}(y+z) -  \int_0^1 \D s\left[ s\,z^i z^j\,\right. \right.\\
&\left.\left.\times \partial_c[\Upsilon(y+[s-1]z, y+sz)\phi^{|c}(y+sz)]  +2\,\Upsilon(y+[s-1]z, y+sz)\,z^{(i}\phi^{|j)}(y+sz) \right.\right.\\
&\left.\left.+ 2\,r\,z^{(i} \partial_4[\Upsilon(y+[s-1]z, y+sz) \dot \phi^{|j)}(y+sz)]\right]\right\}
\end{split}
\end{equation}
that, including (\ref{padic:Ups}), simplifies as
\begin{equation}
\begin{split}
\Theta^{ij}&(\phi,y) = - \Xi(T_y\phi,y) \delta^{ij} - \frac{1}{2} (\mathcal{T}\ast\phi)(y) \int_{\RR^3}\D\mathbf{z}\,\mathcal{G}_3(\mathbf{z})\,z^i \phi^{|j}(\mathbf{y}+\mathbf{z},\tau)\\
& + \frac{1}{2} \int^1_0 \D s\int_{\RR^3}\D\mathbf{z} \,\mathcal{G}_3(\mathbf{z})\left\{ s\,z^i\,z^j \partial_c\left[(\mathcal{T}\ast\phi)(\mathbf{y}+(s-1)\,\mathbf{z}, \tau)\,\phi^{|c}(\mathbf{y}+s\,\mathbf{z}, \tau)\right]\right.\\
&\left.+  2\,(\mathcal{T}\ast\phi)(\mathbf{y}+(s-1)\,\mathbf{z},\tau)\,z^{(i}\phi^{|j)}(\mathbf{y}+s\,\mathbf{z},\tau) +2\,r\,z^{(i}\partial_4\left[(\mathcal{T}\ast\phi)(\mathbf{y}+(s-1)\,\mathbf{z},\tau)\dot\phi^{j)}(\mathbf{y}+s\,\mathbf{z},\tau)\right]\right\}\;.
\end{split}
\end{equation}
Again, taking (\ref{pertsol}), (\ref{tcp}) and (\ref{padic:convh}), the last expression becomes
\begin{equation}
 \Theta^{ij}(y) = \frac{\delta^{ij}}{6} +  \kappa\,\overset{1}{\Theta}\,^{ij}(y) + \mathcal{O}(\kappa^2)
\end{equation} 
where
\begin{equation}
\begin{split}
&\overset{1}{\Theta}\,^{ij}(y) = \frac{1}{2}\int_{\RR^3}\frac{\D\mathbf{k}}{(2\pi)^3}\sum_\nu A_\nu(\mathbf{k})\,e^{i(\alpha_\nu(\mathbf{k})\tau + \mathbf{k}\cdot\mathbf{y})}\left[ \left(\delta^{ij}+2\,r\,k^i\,k^j\right)\,e^{-r|\mathbf{k}|^2}\right.\\
&\qquad \left. + 2\frac{k^i\,k^j}{|\mathbf{k}|^2}\left\{\left(e^{-r|\mathbf{k}|^2}-1\right)\left(1- r\,\alpha_\nu(\mathbf{k})^2\right) + \frac{|\mathbf{k}|^2-\alpha_\nu(\mathbf{k})^2}{|\mathbf{k}|^2}\left(1-\left[1+r\,|\mathbf{k}|^2\right]e^{-r|\mathbf{k}|^2} \right)\right\} \right]\,.
\end{split}
\end{equation}
Again, it is important to highlight this result. Note that the tensor is completely symmetric in $(i,j)$-indices due to the Belinfante-Ronsenfeld symmetrization technique, as expected. Therefore, this fact ensures that we can use this tensor in theories where it is needed to be symmetric, for instance, General Relativity.  Likewise, if we calculate the pressure exerted on a spherical surface $A$ of radius $R$ at $\tau=0$, we get  
\begin{equation}
\Sigma^i:= \int_A \Theta^{ij}(\mathbf{y},0)\,\D^2A_j =  \kappa\:\overset{1}{\Sigma}\,^{i} + \mathcal{O}(\kappa^2)
\end{equation}
where 
\begin{equation}
\begin{split}
\overset{1}{\Sigma}\,^{i} :=&\,4\,\pi\,i\,\int_{\RR^3}\frac{\D\mathbf{k}}{(2\pi)^3}\sum_\nu A_\nu(\mathbf{k})\,\frac{\hat{k}^i}{|\mathbf{k}|^2}\,\left[\sin(R\,|\mathbf{k}|)-|\mathbf{k}|\,R\,\cos(R\,|\mathbf{k}|)\right] \left[ \frac{1}{2}\left(1+2\,r\,|\mathbf{k}|^2\right)e^{-r|\mathbf{k}|^2}\right.\\
&\left. + \left(e^{-r|\mathbf{k}|^2}-1\right)\left(1- r\,\alpha_\nu(\mathbf{k})^2\right) +  \frac{|\mathbf{k}|^2-\alpha_\nu(\mathbf{k})^2}{|\mathbf{k}|^2}\left(1-\left[1+r\,|\mathbf{k}|^2\right]e^{-r|\mathbf{k}|^2} \right)  \right]\,,
\end{split}
\end{equation}
and $\hat{k}^i$ is the unitary vector of $k^i$. One might think that the pressure is imaginary because of the $i$ factor in front, however, it can be proved that, using equations (\ref{padic:alphas}) and (\ref{padic:As}), it is indeed a real value, as expected. 

\vspace{0.1cm}
Making the analogy with the electromagnetic case, the last two remaining elements are the elements of the Poynting vector 
\begin{equation}
\begin{split}
&\Theta^{i4}(\phi,y) = \int_{\RR^4} \D z\left\{\Upsilon(y, y+z)\,z^i\,\dot\phi(y+z) + \int^1_0 \D s\left[\,s\,z^4\,z^i\partial_c\left[\Upsilon(y+[s-1]z, y+sz)\phi^{|c}(y+sz)\right]\right.\right.\\
&\left.\left. + 2\,\Upsilon(y+[s-1]z, y+sz)z^{(4}\,\phi^{|i)}(y+sz)- 2\,r\,z^{(i}\partial_j\left[\Upsilon(y+[s-1]z, y+sz) \dot\phi^{|j)}(y+sz)\right] \right] \right\}
\end{split}
\end{equation}
and
\begin{equation}
\begin{split}
&\Theta^{4i}(\phi,y) = \int_{\RR^4} \D z\left\{ -\Upsilon(y, y+z)\,z^4\,\phi^{|i}(y+z) +2\,r\,\left[\Upsilon(y, y+z)\dot\phi^{|i}(y+z)-\Upsilon(y-z, y)\dot\phi^{|i}(y)\right]\right.\\
&\left. + \int^1_0 \D s\left[\,s\,z^4\,z^i\partial_c\left[\Upsilon(y+[s-1]z, y+sz)\phi^{|c}(y+sz)\right]+ 2\,\Upsilon(y+[s-1]z, y+sz)z^{(4}\,\phi^{|i)}(y+sz)\right.\right.\\
&\left.\left. - 2\,r\,z^{(i}\partial_j\left[\Upsilon(y+[s-1]z, y+sz) \dot\phi^{|j)}(y+sz)\right] \right] \right\}
\end{split}
\end{equation}
respectively. Including (\ref{padic:Ups}) and 
\begin{equation}
\begin{split}
\Upsilon(y, y+z) &= \frac{1}{2}\left[\delta(\xi)\,\mathcal{G}_3(\mathbf{z})\,(\mathcal{T}\ast\phi)(y) - \delta(\mathbf{z})\,\mathcal{G}_1(\xi)\,(\mathcal{E}\ast\phi)(y)\right]\\
\Upsilon(y-z, y) &=\frac{1}{2}\left[\delta(\xi)\,\mathcal{G}_3(\mathbf{z})\,(\mathcal{T}\ast\phi)(y-z) - \delta(\mathbf{z})\,\mathcal{G}_1(\xi)\,(\mathcal{E}\ast\phi)(y-z)\right]\,,
\end{split}
\end{equation}
they become
\begin{equation}
\begin{split}
\Theta^{i4}(\phi,y) =& \frac{1}{2}(\mathcal{T}\ast\phi)(y)\int_{\RR^3}\D\mathbf{z}\,z^i\,\mathcal{G}_3(\mathbf{z})\, \dot\phi(\mathbf{y}+\mathbf{z},\tau)\\
&-\frac{1}{2}\int^1_0\D s\left[\int_\RR \D\xi\,\xi\,\mathcal{G}_1(\xi)(\mathcal{E}\ast\phi)(\mathbf{y},\tau+(s-1)\xi)\,\phi^{|i}(\mathbf{y},\tau+s\,\xi)\right.\\
&\left. + \int_{\RR^3} \D\mathbf{z}\,z^i\,\mathcal{G}_3(\mathbf{z})(\mathcal{T}\ast\phi)(\mathbf{y}+(s-1)\,\mathbf{y},\tau)\,\dot\phi(\mathbf{y}+s\,\mathbf{z},\tau)\right.\\
&\left.  + 2\,r\, \int_{\RR^3}\D\mathbf{z}\,\mathcal{G}_3(\mathbf{z})\,z^{(i}\,\partial_j\left[(\mathcal{T}\ast\phi)(\mathbf{y}+(s-1)\,\mathbf{z},\tau)\,\dot\phi^{|j)}(\mathbf{y}+s\,\mathbf{y},\tau)\right]  \right]
\end{split}
\end{equation}
and
\begin{equation}
\begin{split}
\Theta^{4i}(\phi,y) =& (\mathcal{E}\ast\phi)(y)\int_\RR\D\xi\,\mathcal{G}_1(\xi)\left[\frac{1}{2}\,\xi\, \phi^{|i}(\mathbf{y},\tau+\xi) -r\,\dot\phi^{|i}(\mathbf{y},\tau+\xi)\right]\\
&+ r\,\dot\phi^{|i}(y) \left[\int_\RR\D\xi\,\mathcal{G}_1(\xi)(\mathcal{E}\ast\phi)(\mathbf{y},\tau-\xi) - \int_{\RR^3} \D\mathbf{z}\,\mathcal{G}_3(\mathbf{z})(\mathcal{T}\ast\phi)(\mathbf{y}-\mathbf{z},\tau)\right]\\
& + r\,(\mathcal{T}\ast\phi)(y) \int_{\RR^3}\D\mathbf{z} \,\mathcal{G}_3(\mathbf{z})\,\dot\phi^{|i}(\mathbf{y}+\mathbf{z},\tau)\\
&-\frac{1}{2}\int^1_0\D s\left[\int_\RR \D\xi\,\xi\,\mathcal{G}_1(\xi)(\mathcal{E}\ast\phi)(\mathbf{y},\tau+(s-1)\xi)\,\phi^{|i}(\mathbf{y},\tau+s\,\xi)\right.\\
&\left. + \int_{\RR^3} \D\mathbf{z}\,z^i\,\mathcal{G}_3(\mathbf{z})(\mathcal{T}\ast\phi)(\mathbf{y}+(s-1)\,\mathbf{z},\tau)\,\dot\phi(\mathbf{y}+s\,\mathbf{z},\tau)\right.\\
&\left.  + 2\,r\int_{\RR^3}\D\mathbf{z}\,\mathcal{G}_3(\mathbf{z})\,z^{(i}\,\partial_j\left[(\mathcal{T}\ast\phi)(\mathbf{y}+(s-1)\,\mathbf{z},\tau)\,\dot\phi^{|j)}(\mathbf{y}+s\,\mathbf{y},\tau)\right] \right]\,.
\end{split}
\end{equation}
As above, taking (\ref{pertsol}), (\ref{tcp}) and (\ref{padic:convh}), we finally obtain --- after a tedious computation---
\begin{equation}
 \Theta^{4i}(y) = \Theta^{i4}(y) =  \kappa\:\overset{1}{\Theta}\,^{i4}(y) + \kappa^2\:\overset{2}{\Theta}\,^{i4}(y) + \mathcal{O}(\kappa^3)
\end{equation}  
where 
\begin{equation}
\begin{split}
\overset{1}{\Theta}\,^{i4}(y) = \frac{\kappa}{2}\int_{\RR^3}\frac{\D \mathbf{k}}{(2\pi)^3}& \sum_\nu A_\nu(\mathbf{k}) \:e^{i(\alpha_\nu(\mathbf{k})\tau + \mathbf{k}\cdot\mathbf{y})}\,k^i\\
&\times\left[\frac{\alpha_\nu(\mathbf{k})}{|\mathbf{k}|^2}\left(1-e^{-r|\mathbf{k}|^2}\right) + \frac{1}{\alpha_\nu(\mathbf{k})}\left(1-2\,e^{-r|\mathbf{k}|^2}\right) -2\,r\,\alpha_\nu(\mathbf{k})\right]
\end{split}
\end{equation}
and
\begin{equation}
\begin{split}
&\overset{2}{\Theta}\,^{i4}(y)=\frac{1}{2} \int_{\RR^6}\frac{\D \mathbf{k}\:\D\mathbf{p}}{(2\pi)^6} \sum_{\nu,\mu}A_\nu(\mathbf{k})A_\mu(\mathbf{p})\,e^{i([\alpha_\nu(\mathbf{k})+\alpha_\mu(\mathbf{p})]\tau + [\mathbf{k}+\mathbf{p}]\cdot\mathbf{y})}\\
&\times\left[k^i\left(\frac{e^{-r|\mathbf{p}|^2}}{\alpha_\nu(\mathbf{k})+ \alpha_\mu(\mathbf{k})}\left[e^{-r\alpha_\mu(\mathbf{k})^2} - e^{-r\alpha_\nu(\mathbf{k})^2}\right] - 4\,r\, \alpha_\nu(\mathbf{k})\,e^{-r(|\mathbf{k}|^2+|\mathbf{p}|^2)}\right)\right.\\
&\left. + \frac{(\mathbf{k}+\mathbf{p})^i(\alpha_\nu(\mathbf{k}) + \alpha_\mu(\mathbf{p}))}{f_{\nu\mu}(\mathbf{k},\mathbf{p})}\left(2\,r\,\left[1-2\,e^{-r|\mathbf{k}+\mathbf{p}|^2}\right] + \frac{1-e^{-r(\alpha_\nu(\mathbf{k})+\alpha_\mu(\mathbf{p}))^2}}{(\alpha_\nu(\mathbf{k})+\alpha_\mu(\mathbf{p}))^2} + \frac{e^{-r|\mathbf{k}+\mathbf{p}|^2}-1}{|\mathbf{k}+\mathbf{p}|^2}\right) \right]\,.
\end{split}
\end{equation}
Now, if we integrate the whole volume at $\tau = 0$, we obtain the $i$-components of linear momentum $P^i$, 
\begin{equation}
P^i := \int_{\RR^3}\D\mathbf{y}\,\Theta^{i4}(\mathbf{y},0) = -2\,r\,\kappa^2 \int_{\RR^3}\frac{\D\mathbf{k}}{(2\pi)^3}\sum_{\mu\neq\nu^\prime,\nu} A_\nu(\mathbf{k}) A_{\mu}(-\mathbf{k}) \,k^i\,\alpha_\nu(\mathbf{k})\,e^{-2\,r\,|\mathbf{k}|^2} + \mathcal{O}(\kappa^3)\,.
\end{equation}
With equations (\ref{padic:alphas}) and (\ref{padic:As}), it can be proved that the last expression is real, as expected. Therefore, everything holds.

\section{Conclusion}
We have considered field theories governed by non-local Lagrangians. They differ from local Lagrangian fields usually found in textbooks especially concerning the initial data problem.  In the local (first-order) case, the field equations form a partial differential system with a well-posed initial value problem, and the Cauchy-Kowalevski theorem \cite{john1991} ensures the existence of a solution determined by the field itself and its first time derivative on a non-characteristic hypersurface.
In contrast, the field equations in the non-local case are of integro-differential type ---usually in convolutional form---, and there is no general theorem of existence and uniqueness of solutions for such a system. 

\vspace{0.1cm}
In our approach, as we did in the previous paper on non-local mechanics \cite{Heredia2}, the non-local field equations are taken as constraints selecting the dynamic fields as a subclass $\mathcal{D}$ among all kinematic ones $\mathcal{K}$, and the spacetime evolution is considered as the trivial correspondence $\,\phi(y) \rightarrow \phi(y+x)\,$, i. e. it consists in advancing the ``initial'' spacetime point an amount $x$.

\vspace{0.1cm}
We have then posed the variational problem and derived the Lagrange field equations. It has the peculiarity that, because the non-locality makes all the values of the field intervene in the action integral $S$, the support of the action integral has to be overall $\RR^4$. This fact could lead to the action integral $S$ being infinite. To avoid this issue, the field variations considered are of bounded support. 

\vspace{0.1cm}
We have then generalised Noether's theorem to the case of a non-local Lagrangian and defined the energy density in a closed form. Considering then the structure of the latter, we make an educated guess to define the canonical momenta which we use to set up a Hamiltonian formalism for the non-local field. Notice that this could not be done in the usual manner, i. e. a Legendre transformation consisting of replacing the field time derivatives with the conjugated momenta. 

\vspace{0.1cm}
We start by considering an almost trivial Hamiltonian formalism on the kinematic phase space $\Gamma^\prime$. We then see that the Hamiltonian flow preserves a submanifold that is diffeomorphic to the extended dynamic space $\mathcal{D}^\prime$. This fact enables us to translate the Hamiltonian formalism in the larger space $\Gamma^\prime$ onto $\mathcal{D}^\prime$. We opt for the symplectic formalism instead of Dirac's method \cite{Dirac1964} for constrained Hamiltonian systems since it is better suited by means of pullback techniques. In this way, provided that we are able to find an appropriate coordinatization of the dynamic space, we can derive the formulae for the Hamiltonian and the symplectic form. This fact implies using the field equations as the constraints defining that space, therefore it has to be done specifically for each particular case. 

\vspace{0.1cm}
We have then applied our result to the $p$-adic open string. We have focused on the perturbative solutions allowed by this model to obtain both the Hamiltonian and the symplectic form. Furthermore, the canonical momentum energy tensor and the Belinfante-Rosenfeld tensor were calculated in closed form, and the components of the Belinfante-Rosenfeld tensor were explicitly computed for the perturbative solution. This model has been previously studied in the literature --- see for instance \cite{Moeller2002} --- by other methods. They rely on transforming the non-local Lagrangian into an infinite order Lagrangian by replacing the whole trajectories in the non-local Lagrangian with a formal Taylor series (that includes all the derivatives of the coordinates) and then dealing with it as a higher-order Lagrangian with $n=\infty$. The value of those methods might only be heuristic unless the convergence of the series is proved or the ``convergence'' for $n\rightarrow\infty$ is suitably defined. Furthermore, these methods are cumbersome in that they often imply handling infinite series with many subindices, square $\infty\times\infty$ matrices, formal inverses, regularizations, etc.  In contrast, our approach is based on functional methods and, as it involves integrals instead of series, is much easier to handle.

\vspace{0.1cm}
The same results here obtained could have been derived by converting the non-local Lagrangian into a Lagrangian depending on infinitely many derivatives by means of replacing the field in the non-local Lagrangian with a formal Taylor series. Processing then the infinite-order Lagrangian as if it were a Lagrangian of order $n$ but replacing $n$ with $\infty$, one can obtain an extension of Noether's theorem and, as a consequence of Poincar\'e invariance, derive the energy-momentum and angular momentum tensors in the form of infinite series, which can be summed \cite{Marnelius1973,Jaen1987} to obtain integral expressions over the field derivatives up to a finite order. In ref. \cite{Heredia2020v1},  we have applied this procedure to the non-local Lagrangian of the electromagnetic field in a dispersive medium. We have not taken this path here because, on the one hand, being based on formal, questionable convergence Taylor series, it is a purely heuristic method  and lacks of mathematical rigour and, on the other, it is much more tedious.

\section*{Acknowledgment}
J.LL. was supported by the Spanish MINCIU and ERDF (project ref. RTI2018-098117-B-C22).

\section*{Appendix A}
As far as this article is concerned, we use the following convention for the Fourier transform
\begin{equation}
\tilde g(k) = \mathcal{F}[g](k) =  \int_{\mathbb{R}^n} \D x\:g(x)\,e^{-i k\cdot x} \quad \mathrm{and} \quad g(x) =\mathcal{F}^{-1}[\tilde g](x) = \frac{1}{(2\pi)^n} \int_{\mathbb{R}^n} \D k \:\tilde{g}(k)\,e^{ik\cdot x} \;.
\end{equation}

\subsection*{A.1: Equation (\ref{p6}) derived from (\ref{p6a}) }
Equation (\ref{p6a}) amounts to
$$ \E \ast X = 0\,, \qquad {\rm where} \qquad X:=\T \ast\phi -\left(\E \ast\:\phi\right)^p \,.$$

For any fixed value of $t$, $\phi(\mathbf{x},t)$ belongs to the class $\theta_{M}(\mathbb{R}^3)\,$ of slowly growing smooth functions, i. e. it grows more slowly than any power of $|\mathbf{x}| $ at spatial infinity. This class is a subset of the space of tempered distributions $\mathcal{S}^\prime(\mathbb{R}^3)$ and, as a consequence of this fact, $\, \left(\mathcal{T}\ast \phi\right) (\mathbf{x},t) \,, \left(\mathcal{E}\ast \phi\right) (\mathbf{x},t) \,$ and $X(\mathbf{x},t)$ also belong to $\theta_M(\mathbb{R}^3)$ for any fixed value of $t$ \cite{Vladimirov_GF}.

\vspace{0.1cm}
We will now prove that $ \mathcal{E}\ast X =0 \,$ implies that $X=0$. Indeed, for any fixed $t$, $X_{(t)}(\mathbf{x}) := X(\mathbf{x},t) \,$ is a tempered distribution. Moreover, as $\mathcal{G}_3 \in \mathcal{S}(\mathbb{R}^3)$ --- the space of basic functions ---, the convolution theorem holds \cite{Vladimirov_GF} and 
$$ \mathcal{G}_3 \ast X_{(t)} \in \theta_M(\mathbb{R}^3) \qquad {\rm and} \qquad 
\mathcal{F}\left(\mathcal{G}_3 \ast X_{(t)}\right) = \mathcal{F}\left(\mathcal{G}_3\right) \,\mathcal{F}\left(X_{(t)}\right) \in \mathcal{S}^\prime(\mathbb{R}^3)\,, $$
where $\mathcal{F}$ means the Fourier transform in $\,\mathcal{S}^\prime(\mathbb{R}^3)\,$. 

\vspace{0.1cm}
Therefore, $\,\mathcal{E}\ast X = 0\,$ implies that $\mathcal{G}_3\ast X_{(t)} = 0\,$, whose Fourier transform yields
$$ e^{-r \mathbf{k}^2} \,\mathcal{F}\left(X_{(t)}\right)  = 0 \,,$$
that is $\,\forall \varphi(\mathbf{k})\in \mathcal{S}(\mathbb{R}^3)\,, \qquad \left(\mathcal{F}\left(X_{(t)}\right)\,, e^{-r \mathbf{k}^2}\varphi(\mathbf{k}) \right) = 0 \,$. This fact does not yet imply that $\mathcal{F}\left(X_{(t)}\right) = 0\,$ because not all $\psi\in \mathcal{S}(\mathbb{R}^3)\,$ can be written as $\,e^{-r \mathbf{k}^2}\varphi(\mathbf{k})\,$. However, since $\mathcal{S}^\prime \subset \mathcal{D}^\prime\,$ ---tempered distributions are distributions--- and as, for any $\rho(\mathbf{k})\in \mathcal{D}(\mathbb{R}^3)$, then $\,e^{r\mathbf{k}^2}\,\rho(\mathbf{k}) \,$ also has compact support, we have that
$$ \left(\mathcal{F}\left(X_{(t)}\right),\rho(\mathbf{k})\right) = \left(e^{-r \mathbf{k}^2} \,\mathcal{F}\left(X_{(t)}\right), e^{r \mathbf{k}^2} \,\rho(\mathbf{k})\right) = 0\;.$$
Now, the space of test functions $\mathcal{D}$ is dense in the space of basic functions $\mathcal{S}$ \cite{Vladimirov_GF} , i. e. any $\psi\in \mathcal{S}$ is the limit of a sequence $\,\{\rho_n \in \mathcal{D}\,, \; n\in\mathbb{N}\}\,$ and therefore\\[1ex]
\hspace*{9em} $\displaystyle{ \left(\mathcal{F}\left(X_{(t)}\right),\psi(\mathbf{k})\right) = \lim_{n\rightarrow\infty}\left(\mathcal{F}\left(X_{(t)}\right),\rho_n(\mathbf{k})\right) = 0 \,.}$ \hfill$\Box$

\subsection*{A.2: The convergence of the solution (\ref{p10}) for the $p$-adic open string}
\label{Appendix:C}
The space integral in (\ref{p10}) might diverge because the imaginary part of $\alpha_\nu(\mathbf{k})$ makes that $e^{i\,t\, \alpha_\nu(\mathbf{k})}$ grows exponentially for large $\mathbf{k}$ and $t< 0$.
We will see that, despite this fact, the integral does converge. Indeed, 
\begin{equation}
\label{padic:Desig}
\left|\Phi_0(\mathbf{x},t)\right|=\frac{1}{(2\pi)^3} \left| \intRRR \D\mathbf{k} \sum_{\nu} A_\nu(\mathbf{k}) e^{i[\alpha_\nu(\mathbf{k}) t + \mathbf{k}\cdot\mathbf{x}]}\right| \leq \intRRR \D\mathbf{k} \sum_{\nu}  \left| A_\nu(\mathbf{k})\right| e^{- t\,s\,\alpha_{\nu\mathbb{I}}(\mathbf{k}) }\:,
\end{equation}
where we have written $\alpha_\nu (\mathbf{k})= s\,\left[\alpha_{\nu\mathbb{R}} (\mathbf{k}) + i\,\alpha_{\nu\mathbb{I}}(\mathbf{k}) \right]\,$, with $\;\nu=(s,n)\,, \quad n\in\ZZ$,
\begin{eqnarray*}   
\alpha_{\nu\mathbb{R}} (\mathbf{k}) &:= & \sqrt{\frac{|\mathbf{k}|^2 -2 +\sqrt{\left(|\mathbf{k}|^2-2\right)^2+\left(\frac{2n\pi}{r}\right)^2}}{2}}  \qquad {\rm and} \\[1ex]
\alpha_{\nu\mathbb{I}}(\mathbf{k}) &:= &  \sign(n)\,\sqrt{\frac{-|\mathbf{k}|^2 + 2 +\sqrt{\left(|\mathbf{k}|^2-2\right)^2+\left(\frac{2n\pi}{r}\right)^2}}{2}} \,.
\end{eqnarray*}

If $\,t\,s\,\sign(n) <0\,$, the exponent in the right-hand side of (\ref{padic:Desig}) is $\,|t\,\alpha_{\nu\mathbb{I}}(\mathbf{k})| > 0\,$, which is positive. However, it can be easily checked that $\, \displaystyle{\lim_{|\mathbf{k}|\rightarrow\infty} \alpha_{\nu \mathbb{I}}(\mathbf{k}) = 0 } \,$. Hence,\\
\centerline{$ \forall \varepsilon >0\,$, $\quad\exists K > 0\,\quad$ such that $\quad |\mathbf{k}| > K \quad \Rightarrow\quad |t\,\alpha_{\nu\mathbb{I}}(\mathbf{k})| <\varepsilon $ \,.}
That is, if  $|\mathbf{k}|$ is large enough, the exponential is bounded by the constant $e^\varepsilon$; whence it follows that $\left|\Phi_0(\mathbf{x},t)\right| < \infty$, provided that 
$\displaystyle{ \sum_{\nu} \left| A_\nu(\mathbf{k})\right|} \,$ is summable.

\subsection*{A.3: Derivation of the symplectic form for the $p$-adic open string}
\label{Appendix:CSF}
In this subsection of the appendix, we give the details of the derivation of the symplectic form (\ref{padic:simpletic}) from (\ref{padic:simpletic0}). We take the perturbative solution (\ref{pertsol}) up to second-order in $\kappa$ and obtain that 
\begin{eqnarray*}
\delta\phi(y) &=& \kappa \intRRR \frac{\D\mathbf{k}}{(2\pi)^3} \sum_\nu \delta A_\nu(\mathbf{k})\left\{e^{i[\alpha_\nu(\mathbf{k})\tau+\mathbf{k}\cdot\mathbf{y}]} + \right.  \\[1ex]
 & &\left.+ 2\kappa \intRRR \frac{\D\mathbf{p}}{(2\pi)^3} \sum_{\mu} A_\mu(\mathbf{p}) \frac{e^{i([\alpha_\nu(\mathbf{k})+\alpha_\mu(\mathbf{p})] \tau + [\mathbf{k} + \mathbf{p}]\cdot\mathbf{y})}}{f_{\nu\mu}(\mathbf{k},\mathbf{p})}\right\} + \mathcal{O}(\kappa^3)\:.
\end{eqnarray*}
With the latter, we find that $S_1(\mathbf{y},\tau,s) := \left(\E \ast\:\delta\phi\right)(\mathbf{y},\tau-s) \wedge \delta\phi(\mathbf{y},\tau)$ is
\begin{eqnarray}  
 S_1(\mathbf{y},\tau,s) &=& \kappa^2 \int_{\mathbb{R}^6} \frac{\D\mathbf{k}\,\D\mathbf{q}}{(2\pi)^6} \sum_{\nu,\mu} \delta A_\nu(\mathbf{k})\wedge \delta A_\mu(\mathbf{q}) \,e^{-r|\mathbf{k}|^2} e^{i([\alpha_\nu(\mathbf{k}) + \alpha_\mu(\mathbf{q})]\tau - \alpha_\nu(\mathbf{k}) s + [\mathbf{k}+\mathbf{q}]\cdot\mathbf{y})} \nonumber \\[1ex]
  & &\hspace*{-1em}+ 2\kappa^3 \int_{\mathbb{R}^9} \frac{\D\mathbf{q}\, \D\mathbf{p}\, \D\mathbf{k}}{(2\pi)^9} \sum_{\nu,\mu,\sigma} \delta A_\nu(\mathbf{k}) \wedge \delta A_\mu(\mathbf{q}) A_\sigma(\mathbf{p})\,\left\{ \frac{e^{-r|\mathbf{k}|^2-i \alpha_\nu(\mathbf{k}) s}}{f_{\mu\sigma}(\mathbf{q},\mathbf{p})} + \right.  \nonumber\\[1ex]        	 \label{padic:S1}
  & &\hspace*{-1em}\left.+  \frac{e^{-r|\mathbf{k}+\mathbf{p}|^2-i[\alpha_\nu(\mathbf{k}) + \alpha_\sigma(\mathbf{p})]s}}{f_{\nu\sigma}(\mathbf{k},\mathbf{p})}\right\}  e^{i\left(\left[\alpha_\mu(\mathbf{q})+ \alpha_\nu(\mathbf{k}) + \alpha_\sigma(\mathbf{p})\right] \tau+ [\mathbf{k}+\mathbf{q}+\mathbf{p}]\cdot\mathbf{y}\right)} + \mathcal{O}(\kappa^4)\,.
\end{eqnarray}
Taking into account that $\intRRR \D\mathbf{y}\:e^{i\mathbf{k}\cdot\mathbf{y}} = (2\pi)^3 \delta^{(3)}(\mathbf{k})$, the integration over $\mathbf{y}$ of (\ref{padic:S1}) is
\begin{eqnarray*}
\intR \D \mathbf{y}\:S_1(\mathbf{y},\tau,s) &=& \kappa^2 \intRRR \frac{\D\mathbf{k}}{(2\pi)^3} \sum_{\nu,\mu} \delta A_\nu(\mathbf{k})\wedge \delta A_\mu(-\mathbf{k}) e^{-r|\mathbf{k}|^2 + i[(\alpha_\nu(\mathbf{k}) + \alpha_\mu(\mathbf{k}))\tau - \alpha_\nu(\mathbf{k})s]}\\[1ex]
  & &\hspace*{-1em}+ 2\kappa^3 \int_{\mathbb{R}^6} \frac{\D\mathbf{k}\, \D\mathbf{p}}{(2\pi)^6} \sum_{\nu, \mu, \sigma} \delta A_\nu(\mathbf{k}) \wedge \delta A_\mu(-\mathbf{k}-\mathbf{p}) A_\sigma(\mathbf{p})\left\{ \frac{e^{-r|\mathbf{k}|^2-i \alpha_\nu(\mathbf{k}) s}}{f_{\mu\sigma}(-\mathbf{k}-\mathbf{p},\mathbf{p})} \right.\\[1ex]
 &  &\hspace*{1em}\left. + \frac{e^{-r|\mathbf{k}+\mathbf{p}|^2 -i[\alpha_\nu(\mathbf{k}) + \alpha_\sigma(\mathbf{p})]s}}{f_{\nu \sigma}(\mathbf{k},\mathbf{p})} \right\} e^{i[\alpha_\mu(\mathbf{k}+\mathbf{p})+ \alpha_\nu(\mathbf{k}) + \alpha_\sigma(\mathbf{p})]\tau} + \mathcal{O}(\kappa^4) \,,
\end{eqnarray*}
where we have used that $\alpha_\mu(-\mathbf{k}) = \alpha_{\mu}(\mathbf{k})$. Then, integrating over $\tau$, we obtain 
\begin{eqnarray}
& &\hspace*{-4em}\int^s_0 \D\tau \intRRR \D\mathbf{y}\:S_1(\mathbf{y},\tau,s) =  \kappa^2 \intRRR \frac{\D\mathbf{k}}{(2\pi)^3} \sum_{\nu} \delta A_\nu(\mathbf{k})\wedge \delta {A}_{\nu^\prime}(-\mathbf{k})\:e^{-r|\mathbf{k}|^2}\:s\:e^{-i\alpha_\nu (\mathbf{k}) s}   \nonumber \\[1ex]
& &\hspace*{-2em} + i \kappa^2 \intRRR \frac{\D\mathbf{k}}{(2\pi)^3} \sum_{\mu\neq\nu^\prime} \delta A_\nu(\mathbf{k})\wedge \delta {A}_{\mu}(-\mathbf{k})\frac{e^{-r|\mathbf{k}|^2} \left[ e^{-i s \alpha_\nu(\mathbf{k})}- e^{i s \alpha_\mu(\mathbf{k})}\right]}{ \alpha_\nu(\mathbf{k}) +\alpha_\mu(\mathbf{k})} \nonumber \\[1ex]
& &\hspace*{-2em} + 2 i \kappa^3 \int_{\mathbb{R}^6} \frac{\D\mathbf{p}\, \D\mathbf{k}}{(2\pi)^6} \sum_{\nu,\mu,\sigma} \frac{\delta A_\nu(\mathbf{k}) \wedge \delta {A}_\mu(-\mathbf{k}-\mathbf{p})\,A_\sigma(\mathbf{p})}{\alpha_\nu(\mathbf{k}) + \alpha_\sigma(\mathbf{p}) + \alpha_\mu(\mathbf{k}+\mathbf{p})}\,\times 
\nonumber \\[1ex]    \label{f10}
& &\hspace*{-2em} 
\left[\frac{e^{-r|\mathbf{k}|^2} \,\left( e^{-i s \alpha_\nu(\mathbf{k})} - e^{i s [\alpha_\mu(\mathbf{k+p})+\alpha_\sigma(\mathbf{p})]} \right)}{f_{\mu \sigma}(-\mathbf{k}-\mathbf{p},\mathbf{p})} - 
\frac{e^{-r |\mathbf{k}+\mathbf{p}|^2}\,\,\left( e^{i s \alpha_\mu(\mathbf{k+p})} - e^{-i s [\alpha_\nu(\mathbf{k})+\alpha_\sigma(\mathbf{p})]} \right)}{f_{\nu \sigma}(\mathbf{k},\mathbf{p})} \right]+ \mathcal{O}(\kappa^4) \,, \nonumber
\end{eqnarray}
where we have included that   
\begin{equation}  \label{pS9a}
\int_0^s \D\tau\,e^{i \,a \tau} = \frac{i}{a}\,\left(1- e^{i\,a s}\right) \,, \qquad a\neq 0 \,.
\end{equation}

Finally, substituting (\ref{f10}) into (\ref{padic:simpletic0}), we get
\begin{equation}   \label{pS10}
\omega = 2\,i\,r\,\kappa^2 \intRRR \frac{\D\mathbf{k}}{(2\pi)^3} \sum_\nu \alpha_\nu(\mathbf{k})\,e^{-2r|\mathbf{k}|^2}\,\delta A_\nu(\mathbf{k})\wedge \delta A_{\nu^\prime}(-\mathbf{k}) + \mathcal{O}(\kappa^4) \,,
\end{equation}
where 
we have included that $\alpha_\nu(\mathbf{k})$ is a solution of (\ref{spect1}),
\begin{equation}  \label{f2}
\intR \D s\:\mathcal{G}_1(s)\:e^{-i\alpha_\nu(\mathbf{k}) s}= 2\:e^{-r|\mathbf{k}|^2} \, ,
\qquad  \intR \D s\:\mathcal{G}_1(s)\:s\: e^{-i \alpha_\nu(\mathbf{k})s} = - 4\,i\,r\:\alpha_\nu(\mathbf{k})\:e^{-r|\mathbf{k}|^2}\:.
\end{equation}
and 
$$
\frac{e^{-r |\mathbf{q}|^2}}{f_{\nu \sigma}(\mathbf{k},\mathbf{p})}\:\intR \D s\:\mathcal{G}_1(s)\, \left( e^{-i s [\alpha_\nu(\mathbf{k})+\alpha_\sigma(\mathbf{p})]} - e^{-i s \alpha_\mu(\mathbf{q})} \right) = e^{-r (|\mathbf{q}|^2 + |\mathbf{k}|^2+|\mathbf{p}|^2)}\,.
$$

Now, using that $\,A_{-\nu}(-\mathbf{k}) = \overline{A}_{\nu}(\mathbf{k})\,$ and the fact that $\mathbf{k}$ is a dummy variable, the expression (\ref{pS10}) becomes
$$ \omega = 4\,r\,\kappa^2 \intRRR \frac{\D\mathbf{k}}{(2\pi)^3} \,e^{-2r|\mathbf{k}|^2} \sum_{l\in\mathbb{Z}}\,i\,\alpha_l(\mathbf{k}) \,\delta A_{(+,l)}(\mathbf{k})\wedge \delta \overline{A}_{(+,-l)}(\mathbf{k}) + \mathcal{O}(\kappa^4) \,,$$
where $\,\alpha_l(\mathbf{k}):=\alpha_{(+,l)}(\mathbf{k})\,$. Introducing the new variables 
\begin{equation}  \label{f3}
B_l(\mathbf{k}):= \frac{2 \kappa\,e^{-r|\mathbf{k}|^2}}{(2\pi)^{3/2}} \,\sqrt{r \alpha_l(\mathbf{k})} \, A_{(+,l)}(\mathbf{k}) \qquad {\rm and} \qquad B^\dagger_l(\mathbf{k}):= \overline{B}_{-l}(\mathbf{k}) \,, 
\end{equation}
we finally arrive at
\begin{equation}  \label{f4}
\omega = i\, \intRRR \D\mathbf{k}\,\sum_{l\in\mathbb{Z}} \delta B_l(\mathbf{k})\wedge \delta B^\dagger_l(\mathbf{k}) + \mathcal{O}(\kappa^4) \,.
\end{equation}
We have thus obtained a system of canonical coordinates, namely $\{B_l(\mathbf{k}),B^\dagger_{j}(\mathbf{k}^\prime)\}_{l,j\in\mathbb{Z}}\,$, whose elementary Poisson brackets are 
$$ \{B_l(\mathbf{k}),B^\dagger_{j}(\mathbf{k}^\prime)\} = i\:\delta_{lj}\,\delta(\mathbf{k}-\mathbf{k}^\prime) + \mathcal{O}(\kappa^4)\,, \qquad 
\{B_l(\mathbf{k}), B_{j}(\mathbf{k}^\prime)\} = \{B^\dagger_l(\mathbf{k}), B^\dagger_{j}(\mathbf{k}^\prime)\} =  \mathcal{O}(\kappa^4) \,.$$

\subsection*{Appendix A.4: Derivation of the Hamiltonian for the $p$-adic open string}
Here we show the details of the derivation of the Hamiltonian (\ref{ham1a}) from (\ref{ham1}). Taking the time derivative of the perturbative solution (\ref{pertsol}), we find
\begin{eqnarray}
& &\dot \phi(\mathbf{y},\xi-s) = \frac{i\kappa}{(2\pi)^3}\intR \D\mathbf{k}\sum_{\nu} A_\nu(\mathbf{k})\:\alpha_\nu(\mathbf{k})\: e^{i\left[\alpha_\nu(\mathbf{k})(\xi-s) + \mathbf{k}\cdot\mathbf{y}\right]}   \nonumber\\[1ex]        \label{padic:phip}
& & +\frac{i\kappa^2}{(2\pi)^6} \int_{\mathbb{R}^6} \D\mathbf{k}\, \D\mathbf{p} \sum_{\nu,\mu} \frac{A_\nu(\mathbf{k}) A_\mu(\mathbf{p})}{f_{\nu\mu}(\mathbf{k},\mathbf{p})}\left[\alpha_\nu(\mathbf{k})+\alpha_\mu(\mathbf{p})\right] e^{i[\alpha_\nu(\mathbf{k})+\alpha_\mu(\mathbf{p})](\xi-s) + i (\mathbf{k}+\mathbf{p})\cdot\mathbf{y}} \nonumber\\[1ex]
& &+ \mathcal{O}(\kappa^3)\,.
\end{eqnarray}
Now, the convolutional part in (\ref{ham1}) is
\begin{eqnarray}\label{padic:convh}
& &\left(\E \:\ast\:\phi\right)(\mathbf{y},\xi) = 1 + \frac{\kappa}{(2\pi)^3}\intRRR \D\mathbf{k} \sum_{\nu} A_\nu(\mathbf{k}) e^{-r|\mathbf{k}|^2+ i[\alpha_\nu(\mathbf{k})\xi + \mathbf{k}\cdot\mathbf{y}]}
\nonumber \\[1ex]
 & & +\frac{\kappa^2}{(2\pi)^6}\int_{\mathbb{R}^6}\D\mathbf{k}\,\D\mathbf{p} \sum_{\nu,\mu}\frac{A_\nu(\mathbf{k}) A_\mu(\mathbf{p})}{f_{\nu\mu}(\mathbf{k},\mathbf{p})}e^{-r|\mathbf{k}+\mathbf{p}|^2 + i [\alpha_\nu(\mathbf{k}) + \alpha_\mu(\mathbf{p})] \xi + i (\mathbf{k}+\mathbf{p})\cdot\mathbf{y}}\:.
\end{eqnarray}
Combining both, (\ref{padic:phip}) and (\ref{padic:convh}), and defining $\; S_2(\xi,s) := \intRRR \D\mathbf{y}\:\dot \phi(\mathbf{y},\xi-s) \left(\E \:\ast\:\phi\right)(\mathbf{y},\xi) \,$, we get
\begin{eqnarray}
S_2 (\xi,s) & = & i \kappa \sum_{\nu} \alpha_\nu A_\nu\:e^{i\alpha_\nu(\xi-s)} + i\kappa^2 \intRRR \frac{\D\mathbf{k}}{(2\pi)^3} \sum_{\nu,\mu} A_\nu(\mathbf{k}) \:A_\mu(-\mathbf{k}) \,e^{i [\alpha_\nu(\mathbf{k})+\alpha_\mu(\mathbf{k})]\xi}\times 
\nonumber \\[1ex]   \label{ap-ham1}
 & &  \left\{\frac{\alpha_\nu(\mathbf{k}) +\alpha_\mu(\mathbf{k})}{f_{\nu\mu}(\mathbf{k},-\mathbf{k})}\, e^{-i s [\alpha_\nu(\mathbf{k})+ \alpha_\mu(\mathbf{k})]}  + \alpha_\nu(\mathbf{k}) e^{-r |\mathbf{k}|^2 - i s \alpha_\nu(\mathbf{k})}\right\} \,,
\end{eqnarray}   
where $\alpha_\nu:= \alpha_\nu(0)$, $A_\nu := A_\nu(0)$ and we have used that $\alpha_\nu(-\mathbf{k}) = \alpha_\nu(\mathbf{k})$. Then, integrating over $\xi$, we obtain
\begin{eqnarray*}
\int^s_0 \D\xi\, S_2 (\xi,s) &= & \kappa \sum_{\nu} A_\nu\,\left(1-e^{-i\alpha_\nu s}\right)
+ i \kappa^2 \intRRR \frac{\D\mathbf{k}}{(2\pi)^3} \sum_{\nu} A_\nu(\mathbf{k}) A_{\nu^\prime}(-\mathbf{k})\:s\:\left[\alpha_\nu(\mathbf{k})\:e^{-r |\mathbf{k}|^2 - i s \alpha_\nu(\mathbf{k})} \right. \\[1ex]
 & &\hspace*{-8em} \left. - \frac{e^{-2r |\mathbf{k}|^2}}{r \alpha_\nu(\mathbf{k})}\right] + \kappa^2 \intRRR \frac{\D\mathbf{k}}{(2\pi)^3} \sum_{\mu\neq\nu^\prime} A_\nu(\mathbf{k}) A_{\mu}(-\mathbf{k}) \,
\left[\frac1{f_{\nu\mu} (\mathbf{k},-\mathbf{k})}+ \frac{\alpha_\nu(\mathbf{k})\:e^{-r|\mathbf{k}|^2 + i s \alpha_\mu(\mathbf{k})}}{\alpha_\nu(\mathbf{k}) +\alpha_\mu(\mathbf{k})}\right]\left(1- e^{-i s [\alpha_\nu(\mathbf{k})+\alpha_\mu(\mathbf{k})]} \right)  \,,
\end{eqnarray*}   
where we have included (\ref{pS9a}). Finally, we have that the second term in the right-hand side of (\ref{ham1}) is
\begin{eqnarray}
\frac{1}{2} \intR \D s \,\G_1(s) \int^s_0 \D\xi \,S_2 &=& - \frac{\kappa}{2} \sum_{\nu} A_\nu +\frac{\kappa^2}{(2\pi)^3}\intRRR d\mathbf{k}\left[\sum_{\nu} 2\:r\:\alpha_\nu(\mathbf{k})^2 e^{-2r |\mathbf{k}|^2}A_\nu(\mathbf{k}) \:{A}_\nu(-\mathbf{k}) \right. \nonumber \\[1ex] \label{ham2}
& & \left. - \frac12\, \sum_{\mu\neq \nu} A_\nu(\mathbf{k})  {A}_\mu(-\mathbf{k}) \,\left(\frac1{f_{\nu\mu}(\mathbf{k},-\mathbf{k})} +e^{-2r| \mathbf{k}|^2} \right)\right]  \,,
\end{eqnarray}
where we have used (\ref{f2}).

\vspace{0.1cm}
Now, to calculate the Lagrangian contribution to (\ref{ham1}), we first use (\ref{padic:convh}) to obtain that 
\begin{eqnarray}  \label{tcp}
& & \left(\T\ast\:\phi\right)(y) = 1+ \frac{2\kappa}{(2\pi)^3}\intRRR \D\mathbf{k} \sum_{\nu} A_\nu(\mathbf{k}) e^{-r |\mathbf{k}|^2 + i\left(\alpha_\nu(\mathbf{k}) \tau  +  \mathbf{k}\cdot\mathbf{y}\right)}  \nonumber  \\[1ex]    \label{padic:convt}
& & +\frac{\kappa^2}{(2\pi)^6}\int_{\mathbb{R}^6} \D\mathbf{k}\,\D\mathbf{p} \sum_{\nu,\mu} \frac{A_\nu(\mathbf{k})A_\mu(\mathbf{p})}{f_{\nu\mu}(\mathbf{k},\mathbf{p})} e^{-r[\alpha_\nu(\mathbf{k}) +\alpha_\mu(\mathbf{p})]^2 + i \left([\alpha_\nu(\mathbf{k})+\alpha_\mu(\mathbf{q})]\tau +  (\mathbf{k}+\mathbf{p})\cdot\mathbf{y}\right)}
\end{eqnarray}
and
\begin{eqnarray} \label{ecp3} 
\frac{1}{3}\left[(\E\:\ast\:\phi)(y)\right]^3 &=& \frac{1}{3} + \frac{\kappa}{(2\pi)^3}\intRRR \D\mathbf{k} \sum_{\nu} A_\nu(\mathbf{k}) e^{-r|\mathbf{k}|^2 + i\left(\alpha_\nu(\mathbf{k}) \tau  + \mathbf{k}\cdot\mathbf{y}\right)}   \nonumber \\[1ex]  \label{padic:conv3}
& & \hspace*{-12em}+ \frac{\kappa^2}{(2\pi)^6}\int_{\mathbb{R}^6}\D\mathbf{k}\,\D\mathbf{p} \sum_{\nu,\mu} \frac{A_\nu(\mathbf{k}) A_\mu(\mathbf{p})}{f_{\nu\mu}(\mathbf{k},\mathbf{p})}\,\left[ f_{\nu\mu}(\mathbf{k},\mathbf{p}) + e^{-2 r \mathbf{k}\cdot\mathbf{p}}\right] \, 
e^{-r(|\mathbf{k}|^2 + |\mathbf{p}|^2) + i\left([\alpha_\nu(\mathbf{k}) + \alpha_\mu(\mathbf{p})]\tau + (\mathbf{k}+\mathbf{p})\cdot\mathbf{y}\right)}
\end{eqnarray}   
up to $\kappa^3$ terms. 

\vspace{0.1cm}
Substituting the latter, (\ref{padic:convh}) and  (\ref{padic:convt}) into (\ref{padic:Lntc}), we obtain that
\begin{eqnarray*}
\mathcal{L}(T_y \phi,y)& =& -\frac{1}6 -\frac{\kappa}{2(2\pi)^3} \intRRR \D\mathbf{k} \sum_{\nu} A_\nu(\mathbf{k})\,e^{-r|\mathbf{k}|^2 + i (\alpha_\nu(\mathbf{k}) \tau + \mathbf{k}\cdot\mathbf{y})}  \\[1ex]
& & \hspace*{-5em} - \frac{\kappa^2}{2(2\pi)^6} \int_{\mathbb{R}^6}\D\mathbf{k}\D\mathbf{p} \sum_{\nu,\mu} A_\nu(\mathbf{k}) A_\mu(\mathbf{p}) \, \left[1 +\frac{e^{-2 r \mathbf{k}\cdot\mathbf{p} }}{f_{\nu\mu}(\mathbf{k},\mathbf{p})}\right]\,  e^{-r(|\mathbf{k}|^2+|\mathbf{p}|^2) + i\left([\alpha_\nu(\mathbf{k}) + \alpha_\mu(\mathbf{p})]\tau + (\mathbf{k}+\mathbf{p})\cdot\mathbf{y}\right)}\:,
\end{eqnarray*}
and then, integrating over $\mathbf{y}$ and evaluating at $\tau=0$, the Lagrangian (\ref{ham0}) becomes
\begin{equation}  \label{ham3}
L(\phi) = -\frac{V_y}{6} - \frac{\kappa}{2}\sum_{\nu} A_\nu - \frac{\kappa^2}{2(2\pi)^3}\intRRR \D\mathbf{k} \sum_{\mu \neq\nu^\prime} A_\nu(\mathbf{k}) \,A_{\mu^\prime}(-\mathbf{k}) \,\left[\frac1{f_{\nu\mu}(\mathbf{k},-\mathbf{k})} +e^{-2r|\mathbf{k}|^2}\right] + \mathcal{O}(\kappa^3)\:,
\end{equation}
where $V_y$ stands for $\,\intRRR \D\mathbf{y}$. Finally, by adding together (\ref{ham2}) and (\ref{ham3}), we arrive at the Hamiltonian
$$  h = \frac{V_y}{6} + \frac{\kappa^2}{(2\pi)^3}\intRRR \D\mathbf{k} \sum_{\nu} 2\:r\:\alpha_\nu(\mathbf{k})^2 e^{-2r|\mathbf{k}|^2}A_\nu(\mathbf{k}) A_{\nu^\prime}(-\mathbf{k}) + \mathcal{O}(\kappa^3)\:. $$
Now, using that $\,A_{-\nu}(-\mathbf{k}) = \overline{A}_{\nu}(\mathbf{k})\,$ and the fact that $\mathbf{k}$ is a dummy variable, we easily get that
\begin{equation}  \label{ham4}
h = \frac{V_y}{6} + 4\:r\:\kappa^2 \,\intRRR \frac{\D\mathbf{k}}{(2\pi)^3}  e^{-2r|\mathbf{k}|^2}\,\sum_{l\in\mathbb{Z}} \alpha_l(\mathbf{k})^2\:A_{(+,l)}(\mathbf{k}) \overline{A}_{(+,-l)}(\mathbf{k}) + \mathcal{O}(\kappa^3)
\end{equation}
that, using the variables (\ref{f3}), becomes
\begin{equation}  \label{ham5}
h = \frac{V_y}{6} + \intRRR  \D\mathbf{k}\,\sum_{l\in\mathbb{Z}} \alpha_l(\mathbf{k})\:B_l(\mathbf{k}) B^\dagger_l(\mathbf{k}) + \mathcal{O}(\kappa^3)\:.
\end{equation}

\subsection*{Appendix A.5: The $\phi$ transformation rule for Poincaré invariance.}
Let us find the  transformation rule of $\phi$ to leave the Lagrangian density (\ref{padic:Lntc}) Poincaré invariant. Since $\psi$ is a scalar, by equation (\ref{P3}), we get 
\begin{equation}   \label{A5.1}
 \delta \psi(x) = - \left(\varepsilon^c + \omega^c_{\;b} x^b\right) \,\partial_c\psi(x) \,.
\end{equation}
In order to correctly define the $e^{-r\Box}$ operator, our dynamic variables are the $\phi$ fields. These fields are related to the fields $\psi$ by means of equation (\ref{p2}), which its Fourier transform is
\begin{equation}   \label{A5.2}
\mathcal{F}_{\mathbf{x}}\left[\psi\right](\mathbf{k},t)=\tilde{\mathcal{G}}(\mathbf{k}) \cdot \tilde\phi(k) \,,  \quad\mathrm{with} \quad \tilde\phi(\mathbf{k},t)=\mathcal{F}_{\mathbf{x}}\left[\phi\right](\mathbf{k},t)   \quad \mathrm{and} \quad \tilde{\mathcal{G}}(\mathbf{k}) = e^{-r \mathbf{k}^2}\,.
\end{equation}
On the other hand, the Fourier transform of (\ref{A5.1}) is
\begin{equation}   
\label{A5.3}
\begin{split}
\mathcal{F}_{\mathbf{x}}\left[\delta \psi\right](\mathbf{k},t) & = -\left\{\varepsilon^0\,\partial_0\,\mathcal{F}_{\mathbf{x}}[\psi](\mathbf{k},t) +  \omega^{0i}\,\partial_0\,\mathcal{F}_{\mathbf{x}}[x_i\:\psi](\mathbf{k},t) + (\varepsilon^j +\omega^{j0} x_0) \mathcal{F}_{\mathbf{x}}[\partial_j\psi](\mathbf{k},t) \right.\\
 &\qquad \left. +\omega^{ji}\mathcal{F}_{\mathbf{x}}[x_i\,\partial_j \psi](\mathbf{k},t)  \right\}\,.
 \end{split}
\end{equation}
Plugging equation (\ref{A5.2}) into (\ref{A5.3}), we find 
\begin{equation}   
\label{A5.6}
\begin{split}
 \tilde{\mathcal{G}}(\mathbf{k})\:\delta\tilde\phi(\mathbf{k},t)&= -\tilde{\mathcal{G}}(\mathbf{k})\,\left\{\varepsilon^0\:\partial_0\tilde\phi(\mathbf{k},t) + i\,\omega^{0i}\left[ D_i\partial_0\tilde\phi(\mathbf{k},t)-2\,r\,k_i \,\partial_0\tilde\phi(\mathbf{k},t) \right]\right.\\
 &\quad \left.+ i\,(\varepsilon^j +\omega^{j0} x_0)\mathbf{k}_j \:\tilde\phi(\mathbf{k},t) -\omega^{ji}\left[\delta_{ij} \:\tilde\phi(\mathbf{k},t) + k_j \,D_i\tilde\phi(\mathbf{k},t)\right.\right.\\
 &\quad \left.\left. -2\,r\,k_jk_i \,\tilde\phi(\mathbf{k},t) \right] \right\}
 \end{split}
\end{equation}
where $D_j$ means $\partial/\partial k^j$ to distinguish it from $\partial/\partial x^j$. Bearing mind that $\omega^{ab}=-\omega^{ba}$, we can simplify it as
\begin{equation}   
\label{A5.7}
\begin{split}
 \delta\tilde\phi(\mathbf{k},t)=& - i\,(\varepsilon^j +\omega^{j0} x_0)\,k_j\,\tilde\phi(\mathbf{k},t) - \left(\varepsilon^0 - 2\,i\,r\,\omega^{0i}\,k_i\right)\partial_0\tilde\phi(\mathbf{k},t)\\
 & + \omega^{ji}k_j\,D_i\tilde\phi(\mathbf{k},t) - i\,\omega^{0i}D_i\partial_0\tilde\phi(\mathbf{k},t)
 \end{split}
\end{equation} 
that, undoing the Fourier transform, we obtain how the $\phi$ field transforms,
\begin{equation}
\label{padic:t1}
 \delta \phi(x) = -\left(\varepsilon^c+ \omega^{cb} x_b\right)\partial_c\phi(x) + \omega^{ab}\left[2r\,\delta^4_{[a} \delta^i_{b]}\, \partial_i\dot\phi(x)\right]\,.
\end{equation}

\bibliographystyle{utphys}
{\footnotesize\bibliography{References}}

\end{document}